\documentclass[useAMS,usenatbib]{mn2e}
\usepackage{epsfig}

\title[Particle hydrodynamics with tessellation techniques]
  {Particle hydrodynamics with tessellation techniques}
\author[S.~He{\ss} \& V.~Springel]
  {Steffen~He{\ss}$^1$ and
  Volker~Springel$^1$ \vspace{0.2cm}\\ 
  $^1$Max-Planck-Institut
f\"{u}r Astrophysik, Karl-Schwarzschild-Stra\ss{}e 1, 85740 Garching
bei M\"{u}nchen, Germany}

\renewcommand{\vec}[1]{ {\bmath #1} }

\begin{document}
\label{firstpage}
\maketitle

\begin{abstract}
Lagrangian smoothed particle hydrodynamics (SPH) is a well-established
approach to model fluids in astrophysical problems, thanks to its
geometric flexibility and ability to automatically adjust the spatial
resolution to the clumping of matter.  However, a number of recent
studies have emphasized inaccuracies of SPH in the treatment of fluid
instabilities. The origin of these numerical problems can be traced back
to spurious surface effects across contact discontinuities, and to SPH's
inherent prevention of mixing at the particle level. We here investigate
a new fluid particle model where the density estimate is carried out
with the help of an auxiliary mesh constructed as the Voronoi
tessellation of the simulation particles instead of an adaptive
smoothing kernel. This Voronoi-based approach improves the ability of
the scheme to represent sharp contact discontinuities. We show that this
eliminates spurious surface tension effects present in SPH and that play
a role in suppressing certain fluid instabilities. We find that the new
`Voronoi Particle Hydrodynamics' described here produces {\bf comparable
  results than SPH in shocks, and better ones} in turbulent regimes of
pure hydrodynamical simulations. We also discuss formulations of the
artificial viscosity needed in this scheme and how judiciously chosen
correction forces can be derived in order to maintain a high degree of
particle order and hence a regular Voronoi mesh. This is especially
helpful in simulating self-gravitating fluids with existing gravity
solvers used for N-body simulations.
\end{abstract}


\section{Introduction}

Numerical simulations have become an important research tool in many
areas of astrophysics, in particular in cosmic structure formation and
galaxy formation. This is in part because the physical
conditions involved cannot be reproduced in laboratories on Earth, so that
simulations serve as a replacement for experiments. Perhaps more
importantly, simulations in principle allow a full modeling of all the
involved physics. However, a significant problem in practice is that
that the equations one wants to solve first have to be numerically
discretized  in a suitable fashion. The accuracy of simulations
depends strongly on the properties of this discretization, and it hence
remains an important task to find improved numerical schemes for
astrophysical applications.

In cosmic structure formation, matter is initially essentially 
uniformly distributed, but clusters with time
 under the action of self-gravity to enormous density
contrasts, producing galaxies of vastly different sizes.  Given the
variety of involved geometries, densities and velocities, it is clear
that a Lagrangian method, where the mass of a resolution element stays
(roughly) constant, would be most convenient. This is because a
Lagrangian method automatically concentrates the resolution in regions
where the galaxies form, and hence focuses the numerical effort on the
regions of interest. On the other hand, traditional mesh-based
approaches to hydrodynamics, so-called Eulerian methods, discretize
the volume in a set of cells and do not follow the clustering of
matter, unless this is attempted with a suitable adaptive
mesh-refinement strategy.

The by far most widely used Lagrangian approach in structure formation
is smoothed particle hydrodynamics \citep[SPH, as reviewed
  by][]{Monaghan,Monaghan_2005,Rosswog2009}, a technique that dates
back to particle-based approaches first developed in astronomy more
than 30 years ago \citep{Lucy1977,Gingold_Monaghan,Larson1978}. In
this method the fluid is discretized in terms of particles of fixed
mass, which are used to construct an approximation to the Euler
equations based on the adaptive kernel interpolation technique. SPH
can be very easily coupled to self-gravity, it is remarkably robust
(e.g.~negative densities cannot arise), and the introduction of extra
physics (e.g.~feedback processes in the context of star formation) is
intuitive. All of these properties have made it very popular for
problems such as planet formation or galaxy mergers
\citep[e.g.][]{Mihos,Mayer2002,Disk_Galaxy_Formation,Dolag2005}, where
spatially separated regions of the simulation volume feature widely
different densities.

However, recent studies have highlighted a number of differences in
the results of SPH-based calculations compared to more traditional
grid-based Eulerian methods for hydrodynamics.  For example, the two
methods appear to disagree about the entropy produced in the central
region of a forming galaxy cluster under non-radiative conditions, as
first seen in the `Santa Barbara cluster comparison project'
\citep{Frenk}.  It has been suggested that this problem may be caused
by a suppression of the Raleigh-Taylor fluid instability in SPH
\citep{Mitchell2009} and the lack of mixing at the particle level
\citep{Tasker2008,Wadsley}.  Indeed, \citet{Agertz} have shown that
SPH tends to suppress Kelvin-Helmholtz fluid instabilities in shear
flows across interfaces with sizable density jumps.  In such a
situation, SPH's density estimate leads to spurious forces at the
interface which produce an artificial `gap' in the particle
distribution and a surface tension effect that ultimately produces
errors in the hydrodynamical evolution. To what extent these numerical
artifacts negatively affect the global accuracy of simulations in
practice is unclear, and this can in any case be expected to be
problem dependent. However, an improvement of standard SPH that avoids
these errors is obviously desirable.

First proposals in this direction have recently been made.
\citet{Price2008} suggests to introduce artificial heat conduction into
SPH such that discontinuities in the temperature field are smoothed out,
in analogy to the ordinary artificial viscosity that effectively
smoothes out discontinuities in the velocity field occurring at shocks.
This heat conduction produces a soft instead of an abrupt transition of
the specific entropy across a contact discontinuity, which in turn helps
to better represent the growth of Kelvin-Helmholtz instabilities at such
interfaces. More recently, \citet{Read2009} have modified an idea by
\citet{Ritchie2001} for a modified SPH density estimate that assumes
that the local neighbours have similar pressures, and which is designed
to avoid the `pressure blip' in the standard approach at contact
discontinuities. Together with a modified kernel shape and a drastically
enlarged number of neighbours (by a factor of $\sim 10$, implying a
similar increase in the computational cost), \citet{Read2009} obtained
better growth of Kelvin-Helmholtz instabilities across
density jumps.

In this work we follow a different approach that eliminates the
ordinary SPH kernel altogether. Instead, we use the distribution of
points with variable masses to construct an auxiliary mesh, which is
then used to derive local density estimates. If the particle
hydrodynamics is derived from a Lagrangian, it turns out that
obtaining this density estimate is already sufficient to uniquely
determine the equations of motion.  The use of Delaunay tessellations
to construct density fields from arbitrary point sets has been
discussed in the literature \citep{Schaap,Icke,Weygaert,Pelupessy},
but as we show in this paper, its topological dual, the Voronoi
tessellation, is actually preferable for our hydrodynamical
application. In the Voronoi tessellation, to every particle a
polyhedra is assigned which encompasses the space closer to this
particle than to any other.  Based on these volumes associated with
each particle, local densities and hydrodynamical forces can be
estimated, leading to an interesting alternative to SPH. In
particular, it is immediately clear that unlike SPH this approach
yields a consistent discretization not only of the mass but also of
the volume, which should help to yield an improved representation of
contact discontinuities. We note that a conceptionally similar
approach to Voronoi based particle hydrodynamics was first discussed
by \citet{Serrano} in the context of a mesoscopic fluid particle
model. We here extend this idea to the treatment of the Euler
equations in astrophysical systems.

We emphasize that the method we introduce in this study is radically
different from to the one implemented in the new {\small AREPO} code
\citep{AREPO}.  Whereas the latter is also based on a (moving) Voronoi
tessellation, it employs a finite volume scheme with a Riemann solver
to compute hydrodynamical fluxes across mesh boundaries. This involves
an explicit second-order reconstruction of the fluid throughout the
volume, and allows for changes of the mass contained in each cell even
if the mesh is on average moving with the flow. In contrast, we here
derive a fluid particle model from a discretized Lagrangian in which
the masses of each element stay strictly constant, and in which the
motion of the particles is governed by pairwise pressure force
exchanged between them. While {\small AREPO} is conceptually close to
the techniques used in Eulerian hydrodynamics, the method we study
here is conceptually close to SPH.

This paper is structured as follows. In Section~\ref{SecBasics}, we
discuss how the equations of motion can be derived for the Lagrangian
particle approach to hydrodynamics discussed here. We will also
present suitable formulations of artificial viscosity for our scheme.
In Section~\ref{SecRegularity}, we discuss the role of the regularity
of Voronoi cells and means to improve it. We briefly describe the
implementation of our numerical scheme in a modified version of the
{\small GADGET} code in Section~\ref{SecImplementation}, and then turn
in Section~\ref{SecResults} to a description of results for a
suite of test problems with our new `Voronoi Particle Hydrodynamics'
(VPH) scheme. These tests range from simple shock-tube problems, to
fluid instabilities, and three-dimensional stripping of gas in a
supersonic flow. Finally, we summarize our conclusions in
Section~\ref{SecConclusions}. In two Appendices, we discuss gradient
operators for Voronoi meshes and give the derivation of correction
forces that can be used to maintain very regular mesh geometries, if
desired.

\section{Particle based hydrodynamics} \label{SecBasics}

We begin by introducing our methodology for a particle-based fluid
dynamics based on Voronoi tessellations. This method is close in spirit
to SPH, but differs in important aspects. Where appropriate, we discuss
these differences in detail.

\subsection{A Lagrangian approach for particle based fluid dynamics} \label{A Lagrangian Approach}

We discretize the fluid in terms of $N$ mass elements of mass $m_i$.
The discretized fluid Lagrangian can then be adopted as
\begin{equation}
	L = \sum_i \left[ \: \frac{1}{2}\: m_i \: \vec{v}_i^2 - m_i \:
          u_i(\rho_i, s_i) \: \right] .
\label{lagr}
\end{equation}
This is simply the difference of the kinetic and thermal energy of the
particles. The thermal energy $u_i$ per unit mass depends both on the
density $\rho_i$ and the specific entropy $s_i$ of the particle. In this
work, we aim to approximate inviscid ideal gases, hence the equation of
state (EOS) is that of a polytropic gas, where the pressure is $P_i =
s_i \rho^\gamma$, and the entropic function $s_i$ (or simply `entropy'
for short, since it depends only on the thermodynamic entropy)
labels the adiabat on which this gas element resides.

When the specific entropy $s_i$ and the mass $m_i$ remain constant
for a fluid element $i$, the internal energy $u_i$ changes according to
$\frac{\partial u}{\partial \rho} = \frac{P}{\rho^2}$. Using this
result, we can readily write down the Lagrangian equations of motion of
(reversible) fluid dynamics:
\begin{eqnarray}
m_i \ddot \vec{r}_i	& = & - \sum_j^{N}  m_j \frac{\partial
  u_j}{\partial \vec{r}_i}  \nonumber \\
		& = & - \sum_j^{N}  m_j \frac{\partial u_j}{\partial
  \rho_j} \frac{\partial \rho_j}{\partial \vec{r}_i} \nonumber \\ 
		& = & - \sum_j^{N}  m_j \frac{P_j}{\rho_j^2} \frac{\partial \rho_j}{\partial V_j} \frac{\partial V_j}{\partial \vec{r}_i}
 \label{base} 
\end{eqnarray} 
We see that the primary input required for a more explicit form of the
equations is a density estimate based on the particle coordinates, or
alternatively, an estimate of the volume associated with a given
particle.

SPH addresses this task with a kernel estimation technique to obtain
the density, where an adaptive spherically symmetric smoothing kernel
is employed to calculate the density based on the spatial distribution
of an approximately fixed number of nearest neighbours.  The
Lagrangian then uniquely determines the equations of motion that
simultaneously conserve energy and entropy
\citep{springel_hernquist_02}. However, we note that SPH does not
achieve a consistent volume estimate, i.e.~the sum of the effective
volumes of the particles, $V_i = m_i/\rho_i$, is not guaranteed to be
equal to the total simulated volume. Furthermore, the inherent
smoothing operation in the density estimate is bound to be inaccurate
at contact discontinuities and phase interfaces, where the density may
discontinuously jump by a large factor. In the following, we therefore
look for alternative ways to construct density estimates which improve
on these deficits.

\subsection{Density estimates with tessellation techniques}

One promising approach for more accurate density and specific volume
estimates lies in the use of an auxiliary mesh that is generated by the
particle distribution. A mesh can readily yield a partitioning of the
volume such that the total volume is conserved, and also allows multiple
ways to `spread out' the particle masses $m_i$ in a conservative fashion
such that an estimate of the density field is obtained.

There are two basic geometric constructions that suggest themselves as
such mesh candidates. These are the Delaunay \citep{Dirichlet_} and the
Voronoi tessellations \citep{Voronoj_}, which are in fact mathematically
closely related, as we discuss below.  In the Voronoi tessellation, space
is subdivided into non-overlapping polyhedra which each encompass the
volume which is closer to its corresponding point than to any other
point. The surfaces of these polyhedra are therefore the
bisectors to the nearest neighbours.  The Delaunay tessellation on the
other hand decomposes space into a set of tetrahedra (or triangles in
2D), with vertices at the point coordinates.  The defining property of
the Delaunay tessellation is that the circumcircles of the
tetrahedra do not contain any of the points in their interior.  This
property in fact makes this tessellation uniquely determined for points
in general position.

It turns out that these two tessellations are {\em dual} to each other;
to each edge of the Delaunay tessellation corresponds a face of the
Voronoi tessellation, and the circumcircle centres of the Delaunay
tetrahedra are the vertices of the Voronoi faces. One implication of
this is that the Delaunay and Voronoi Tessellations can be easily
transformed into each other. In practice it is typically
simpler to always construct the Delaunay tessellation, even if one works
with the Voronoi, because the former has more efficient and simpler
algorithms for construction.

Both tessellations can in principle be used to derive density
estimates.  \citet{Schaap2000} introduced the Delaunay Tessellation
Field Estimator (DTFE) technique, and \citet{Pelupessy} showed that it
offers superior resolution compared to SPH-like density estimates for
detecting cosmological large-scale structure \citep[for a review
  see][]{Weygaert2009}. In this approach, the total volume of the
contiguous set of Delaunay cells around a point is used to assign
particle densities, and a full density field can be constructed by
linearly interpolating the densities inside each Delaunay tetrahedron.
As a possible application of this density estimate, \citet{Pelupessy}
also suggested its use in a particle-based hydrodynamic scheme.
However, we caution that a rather serious short-coming of the Delaunay
tessellation in this context is that the tessellation may occasionally
change {\em discontinuously} as a function of the particle
coordinates. This happens whenever a particle moves over the
circumcircle of one of the tetrahedra. An infinitesimal particle
motion can hence be sufficient to create finite changes in the volume
of its associated contiguous Delaunay cell (this is the union of all
Delaunay tetrahedra of which the given point is one of the vertices)
of a particle. As a result, the thermal energy of the point set is not
a continuous function of the particle coordinates.  This makes the
DTFE technique ill suited to be the basis of a hydrodynamical particle
method.

On the other hand, the volumes of the Voronoi cells always depend
continuously on the particle coordinates, despite the fact that
topological changes of the tessellation may occur as a result of
particle motion. This is because flips of edges in the Delaunay
tessellation happen precisely when the corresponding Voronoi faces
have vanishing area. Another advantage of the Voronoi tessellation is
that it is {\em always} uniquely defined for any distribution of the
points, whereas for certain degenerate point sets (those where more
than four points lie on a common circumsphere), more than one valid
Delaunay tessellation may exist, which can then make Delaunay-derived
density estimates non-unique.  We remark that the uniqueness of the
Voronoi tessellation does not hold in reverse, i.e.~a given Voronoi
tessellation can in general be produced by a number of different point
distributions. This has important consequences for the stability of
the scheme, as we discuss later on in more detail.

Based on the above, the Voronoi tessellation is a promising
construction for a particle fluid model, hence we adopt it in the
following. In particular, we shall associate the volume $V_i$ of a
Voronoi cell with its corresponding point, yielding a consistent
decomposition of the total simulated volume. The simplest possible
density estimate is then simply given by
\begin{equation}
\rho_i = \frac{m_i}{V_i}
\end{equation}
which we shall use in this paper. More involved higher-order density field
reconstructions could be considered as well, an idea we leave for future
work.

Based on this density estimate, and given the specific entropies of each
particle, the local pressure and the thermal pressure per unit mass can
be computed. Also, one may define a gradient operator for the Voronoi
mesh (see our discussion in Appendix \ref{app_grad}), which could be used to estimate
pressure gradients, and hence to yield discretizations of the Euler
equations.  However, a better approach is to start from the discretized
Lagrangian, as this automatically gives equations of motion that satisfy
the conservation laws. We shall adopt this strategy in the following.

\subsection{Equations of motion for Voronoi-based particle hydrodynamics}

\begin{figure}
\begin{center}
  \includegraphics[width=0.40\textwidth]{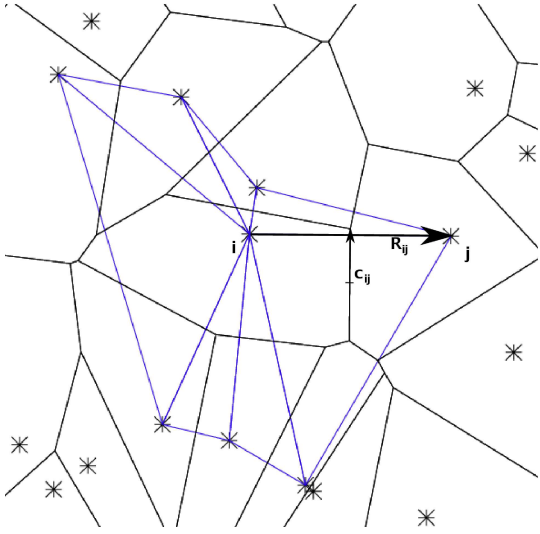}

\caption{Section of a Voronoi diagram for a set of points marked with
  asterisks. All the triangles in the dual Delaunay tessellation that
  are shared by the point in the centre are marked in blue. The vectors
  $\vec{c}_{ij}$ and $\vec{e}_{ij}$ needed to calculate the derivative
  of the volume are also marked. \label{FigVoronoiSketch}}
\end{center}
\end{figure}
 
Since the volumes of the Voronoi cells depend only on the configuration
of the points, we can readily obtain the equations of motion if we find
the partial derivative of a cell volume with respect to any of the
particle coordinates.  We here adopt the result of \citet{Serrano}, who
showed that the relevant derivative is given by \citep[see also][]{De_Fabritiis}
\begin{equation}
 \frac{\partial V_j}{\partial
   \vec{r}_i}=-A_{ij}\left(\frac{\vec{c}_{ij}}{R_{ij}}+\frac{\vec{e}_{ij}}{2}\right)
\;\;\; \mbox{for $i\ne j$,}
\label{Serrano_Espanol}
\end{equation}
where $\frac{\partial }{\partial \vec{r}_i}$ denotes the gradient
operator with respect to $\vec{r}_i$. Here $R_{ij}$ is the distance
between two neighboring points, $\vec{e}_{ij}= (\vec{r}_j-\vec{r}_i)/
R_{ij}$ denotes a unit vector from $i$ to the neighbour $j$, which is
normal to the Voronoi face of area $A_{ij}$ between cells $i$ and $j$.
Formally, we can define $A_{ij}$ for any pair of different particles,
but if $i$ and $j$ are not neighbours in the Voronoi tessellation
(i.e.~do not share a face), we set $A_{ij}\equiv0$, implying that  
in sums that involve the factor $A_{ij}$ only the direct neighbours contribute.
Note that equation (\ref{Serrano_Espanol}) holds only for $j \neq
i$. But one can readily derive an expression for $\partial V_i/\partial
\vec{r}_i$ by invoking volume conservation. This yields
\begin{equation}
 \frac{\partial V_i}{\partial
   \vec{r}_i}= - \sum_{j\neq i} \frac{\partial V_j}{\partial
   \vec{r}_i}.
\label{Serrano_Espanol_NN}
\end{equation}
As sketched in Figure~\ref{FigVoronoiSketch}, the vector
$\vec{c}_{ij}$ points from the midpoint between $i$ and $j$ to the
centroid of the face $A_{ij}$, and is orthogonal to $\vec{e}_{ij}$.
The term involving ${\vec{e}_{ij}}/{2}$ can be easily understood
geometrically from the change of the volumes of the pair of pyramids
spanned by the face between $i$ and $j$ and the two points. But if the
center of the face is displaced from the line connecting $i$ and $j$,
a second term involving $\vec{c}_{ij}$ appears that stems from the
turning of the face when the points are moved.

We are now in a position to write down the resulting equations of
motion, based on equations (\ref{base}), (\ref{Serrano_Espanol}) and
(\ref{Serrano_Espanol_NN}). This first yields
\begin{eqnarray}
\frac{\partial \rho_j}{\partial \vec{r}_i} & = &
  \frac{m_j}{V_j^2} \left[  (1-\delta_{ij})A_{ij}
  \left(\frac{\vec{c}_{ij}}{R_{ij}}+\frac{\vec{e}_{ij}}{2}\right)\right. \nonumber \\ && -
  \delta_{ij}  \sum_{k\neq j} A_{jk} \left. \left(\frac{\vec{c}_{jk}}{R_{jk}} +
  \frac{\vec{e}_{jk}}{2}\right)\right] ,  \label{antisymm}
\end{eqnarray}
which then gives rise to the equations of motion in the form
\begin{equation}
m_i \ddot \vec{r}_i = 
  \sum_{j\neq i}  A_{ij} (P_i-P_j) \left (\frac{\vec{c}_{ij}}{R_{ij}} +
  \frac{\vec{e}_{ij}}{2}\right) .  \label{eqnmotion1}
\end{equation}
This is a rather intuitive
result, as it shows that motions are generated by the pressure
differences that occur across faces of the tessellation. If the
pressures are all equal, the forces vanish exactly, unlike in ordinary
SPH.

In the form of equation (\ref{eqnmotion1}), it is  not obvious
whether the forces between a given pair 
of particles are antisymmetric.
However, noting the identity $\sum_{j\ne i} A_{ij}\vec{e}_{ij} = 0$,
which follows from Gauss' theorem, we can restore
manifest antisymmetry  in the equations of
motion, which is in general preferable for numerical reasons. To this
end, we simply subtract  $P_i \sum_{j\ne i} A_{ij}\vec{e}_{ij} = 0$
from (\ref{eqnmotion1}), yielding our final equations of motion as
\begin{equation}
m_i \ddot \vec{r}_i = 
 - \sum_{j\neq i}  A_{ij}\left[ (P_i+P_j) \frac{\vec{e}_{ij}}{2} +
(P_j - P_i) \frac{\vec{c}_{ij}}{R_{ij}} \right], \label{eqnmotion2}
\end{equation}
which is now pairwise antisymmetric.  Note that whereas formally the
sums appearing in these equations are carried out over all particles,
only the direct neighbours actually contribute, and these are known
from the tessellation. In fact, the list of interacting particle pairs
is exactly given by the list of edges of the underlying Delaunay
tessellation, or equivalently, by the list of faces of the Voronoi
tessellation.

We further note that since the equations of motion have been derived from the
Lagrangian given in equation (\ref{lagr}), these equations conserve
energy, momentum and entropy exactly. In the present form they are hence
a description of the reversible, adiabatic parts of a flow, but they do
not yet contain any dissipation, which is however needed to treat
shocks. If no such dissipation is included, shocks will lead to
unphysical ringing and oscillations in the fluid.

\subsection{Artificial viscosity} \label{Artificial_viscosity}

We follow the standard SPH approach
\citep[e.g.][]{Gingold_Monaghan,Balsara1995} and invoke an artificial
dissipation in the form of an extra friction force that reduces the
kinetic energy and transforms it into heat.  There is great freedom in
the form of this viscous force, but ideally it should only become active
where it is really needed, i.e. in shocks, and should be negligible away
from shocks, such that inviscid behavior is ensured there.  The most
widely used and tested formulation of the viscous acceleration in SPH schemes is
given by
\begin{eqnarray}
	 (\vec{a}_{\mathrm{visc}})_i &=& -  \sum_j 
  m_j \Pi_{ij}  \vec{\nabla}_i \overline{W}_{ij}, \\
    \Pi_{ij} &=&  
  \frac{ \bar{f}_{ij}}{\bar{\rho}_{ij}} \left( -\alpha \, \mu_{ij} \bar{c}_{ij} +
  \beta \, \mu_{ij}^2 \right),  \\
 \mu_{ij} &=&
  \frac{\bar{h}_{ij}\, \vec{v}_{ij} \cdot \vec{r}_{ij}}{r_{ij}^2 + \epsilon\, \bar{h}_{ij}^2} , \\ \bar{f}_{ij} &=&
  \frac{f_i + f_j}{2} \: , \quad f_i \:=\:\frac{|\vec{\nabla} \cdot
    \vec{v}|_i}{|\vec{\nabla} \cdot \vec{v}|_i+|\vec{\nabla} \times \vec{v}|_i +
    \epsilon}, \label{BalsaraF}
\end{eqnarray}
provided that $\vec{v}_{ik} \cdot \vec{r}_{ik} < 0$, i.e.~the
neighboring particles approach each other, otherwise the viscous force
that is mediated by the viscous tensor $\Pi_{ij}$ is set to zero.  In
this notation, $q_{ij}$ represents the difference and $\bar{q}_{ij}$
the average between the quantities $q$ associated with particles $i$
and $j$. The parameter $\epsilon$ is a tiny value introduced to guard
against numerical divergences. The parameters $\alpha$ and $\beta$ set
the strength of the viscosity and are typically set to of order $\sim
1$. The factors $f_i$ measure the strength of the local velocity
dispersion relative to the local shear, and are introduced as
so-called Balsara switch to reduce the viscosity if the local flow is
dominated by shear \citep{Balsara1995}.

The above formulation of a viscous force can be adopted to the Voronoi
scheme in a number of ways. We first define the projected pairwise 
velocity as 
\begin{equation}
  w_{ij} = \frac{\vec{v}_{ij} \cdot \vec{r}_{ij}}{|\vec{r}_{ij}|} ,
\end{equation}
and make the replacement $\mu_{ij} \to w_{ij}$. This is effectively
yielding the `signal velocity' form of the standard viscosity
\citep{Monaghan1997}. For simplicity, we shall also adopt the common
choice $\beta=2\alpha$.  We next recognize that in SPH the viscous
tensor is introduced into the equations of motion as if it was an
extra pressure of the form $P_{\rm visc} = \frac{1}{2} \rho_{ij}^2
\Pi_{ij}$ \citep{gadget2}. Using this analogy, and guiding
ourselves by the form of the Voronoi-based equations of motion (\ref{eqnmotion2}),
we can readily write down a parameterization of the viscous force
acting on a particle as
\begin{equation}
m_i(\vec{a}_{\mathrm{visc}})_i  = 
 - \sum_j  A_{ij} \overline{\rho}_{ij}^2 \Pi_{ij} \frac{\vec{e}_{ij}}{2} .
\label{art_visc1}
\end{equation}
Here we have only introduced a viscous force component parallel to the
line connecting the two particles, since we assume that the `viscous
extra pressure' is the same for a pair of interacting particles,
i.e. $(P_{\rm visc})_i = (P_{\rm visc})_j$. 
A more explicit form of
the viscous acceleration is given by the following expression:
\begin{equation}
(\vec{a}_{\mathrm{visc}})_i = \alpha \sum_j \frac{\bar{f_{ij}}}{m_i}
  \bar{\rho}_{ij} A_{ij} \left( w_{ij} \bar{c}_{ij} - 2 w_{ij}^2
  \right) \frac{\vec{e}_{ij}}{2}.
	\label{art_visc}
\end{equation}
Note that the viscous force is pairwise antisymmetric, and will only
become active if two particles approach each other.  We also want to
stress that artificial viscosity parameterizations different from that
of equation (\ref{art_visc}) are of course possible. We here simply
adopt this form as a first best guess, based on the analogy with the
widely tested SPH formulation.

It is interesting to compare the artificial viscosity with the viscosity
terms of the Navier Stokes equation,
\begin{eqnarray}
 m \frac{D \vec{v}}{D t} &=& - \vec{\nabla} P + \eta \: \Delta \vec{v} +
 \lambda \: \vec{\nabla} (\vec{\nabla} \vec{v}).
\end{eqnarray}
Neglecting the shear viscosity $\eta$ and approximating the gradient
operator with its Voronoi discretized form (see Appendix A),
this becomes
\begin{equation}
 \left(\lambda \: \vec{\nabla} (\vec{\nabla} \vec{v})\right)_i 
 \simeq 
 \lambda \: \frac{1}{V_i} \sum_j A_{ij}
 \frac{\vec{e}_{ij}}{2}
 (\vec{\nabla} \vec{v})_j  .
\end{equation}
Additionally approximating
\begin{eqnarray}
(\nabla \vec{v})_j \approx \frac{\vec{v}_{ij} \cdot
    \vec{r}_{ij}}{|\vec{r}_{ij}|} = w_{ij}
\end{eqnarray}
yields a term like the one linear in $w$ in equation
(\ref{art_visc}), adding some further justification to this form of
the viscous force, which has the form of an artificial bulk viscosity.

In order to maintain energy conservation, heat must be produced at a
rate ${\rm d}E/{\rm d}t$ that exactly balances the loss of kinetic
energy due to the extra friction from the artificial viscosity.  We
inject this energy symmetrically into the specific entropies of the two
particles. Defining the pairwise viscous forces as
\begin{equation}
(\vec{f}_{\rm visc})_{ij} = -   A_{ij} \overline{\rho}_{ij}^2 \Pi_{ij} \frac{\vec{e}_{ij}}{2}, 
\end{equation}
the heating is given by
\begin{equation}
\frac{\mathrm{d} u_i}{\mathrm{d} t} = \frac{1}{2} 
\sum_j (\vec{f}_{\rm visc})_{ij} \cdot \vec{v}_{ij} .
\end{equation}
With $u_i= s_i \rho_i^{\gamma-1} / (\gamma -1) $ this yields for the
rate of entropy production
\begin{equation}
\frac{ {\rm d} s_i} { {\rm d} t}  =  
 \frac{\gamma -1} {2\, \rho_i^{\gamma-1}}
\sum_j  (\vec{f}_{\rm visc})_{ij} \cdot  \vec{v}_{ij} .
\end{equation}
 In this form, the equations still conserve
total energy and momentum, while the change of the total entropy is
positive definite.

The artificial viscosity is necessary to capture shocks and to damp
postshock oscillations in the vicinity of shocks, but everywhere else in
the fluid it can induce spurious dissipation that distorts the physics
of an inviscid gas. In order to reduce the influence of the viscosity in
regions away from shocks, the prefactor $\alpha$ that sets the strength
of the viscosity can be chosen adaptively \citep{Morris,Dolag2005}. The idea 
{\bf of this dynamic viscosity} is
that every particle gets an individual viscosity strength $\alpha$ which
is evolved in time according to the differential equation
\begin{equation}
 \frac{\mathrm{d} \alpha}{\mathrm{d} t}  =  -  \frac{\alpha - \alpha^*}{\tau} \: + S  .
\label{dyn_visc_eq}
\end{equation}
Here $\alpha$ is decaying to a minimum $\alpha^*$ on a timescale $\tau$,
and is increased by the source term $S$. One possible choice for this
source
term is
\begin{equation}
 S  =  \bar{f}_{ij} \: \zeta \: \vec{\nabla} \cdot \vec{v} ,
\end{equation}
which we adopt in our implementation of a time-variable artificial
viscosity, using the discretized estimate of the divergence described
in Appendix~A. Here both the response coefficient $\zeta$ and the
timescale $\tau$ have to be calibrated empirically.  When a shock
arrives in an unperturbed area, $\alpha$ is at its minimum and needs
to jump very quickly to a higher level in order to capture the shock
and prevent post shock oscillations, whereas behind the shock, the
viscosity should quickly return to a low value. However, a too large
value for $\zeta$ may trigger high viscosity  due to the often
noisy estimates of the $\vec{\nabla} \cdot \vec{v}$ term, and if
$\tau$ is too small, the viscosity may decay too quickly to capture
the shock properly. Finally, the minimum viscosity $\alpha^*$ can be
set to a non-zero value to improve particle order and thereby reduce
noise, at the cost of introducing some minimum viscosity.

\subsection{Treatment of mixing} \label{Mixing_Musings}

Hydrodynamic simulations are able to follow the advection of fluids only
down to the resolution scale. But especially Lagrangian schemes do not
include mixing processes of the fluid on sub-resolution scales. In
Eulerian codes, such mixing is implicit whenever a new averaged
thermodynamic state for a cell is computed after fluxes of gas have
entered or left it. This mixing keeps the total energy fixed, but will
in general raise the entropy of the system. In the Lagrangian particle
approach of SPH and in the Voronoi approach developed here, such mixing
effects are, however, entirely suppressed. The specific entropies of
neighbouring particles stay constant, except when a shock is present.
While this reliably eliminates unwanted entropy production from
advection errors, it also prevents the proper subresolution production
of entropy when small-scale fluid instabilities should mix the fluid on
the resolution scale and produce  homogeneous thermodynamic
properties.

We have therefore tried to model this subgrid mixing with a heuristic
model which conjectures that small-scale fluid instabilities, if
present, equalize the local temperature field by mixing. This will
then smooth out sharp contact discontinuities and also tend to
equilibrate the specific entropies of the cells.  Similar ideas have
recently been discussed by \citet{Price2008}, \citet{Wadsley} and
\citet{Shen} in the context of SPH, but our approach differs in
detail.  In particular, we restrict the averaging to shearing layers,
and motivate the timescale for mixing directly with the growth
timescale of the Kelvin-Helmholtz instability on the resolution scale.

The linear theory growth timescale of a perturbation across a contact
discontinuity with densities $\rho_i$ and $\rho_j$ that 
exhibits a jump in the
tangential velocity of size $v_{ij}^{\parallel}$ (a shear layer) is given by
\begin{equation}
 t_{\rm KH} =  \frac{\rho_i + \rho_j}{2\, k\, v_{ij}^{\parallel} \sqrt{\rho_i  \rho_j}}.
\end{equation}
Here $k=2\pi / L$ is the wavenumber of the Kelvin-Helmholtz mode. We
shall assume that $L$ is of order the cell dimension, which is in turn
of order the particle separation, i.e.~we will set $L = R_{ij}$ when a
particle pair of separation $R_{ij}$ is considered. Similarly, the
relevant velocity jump 
is simply the velocity difference projected onto the face between two
particles, which is normal to their separation vector, hence
\begin{equation}
v_{ij}^{\parallel} = |\vec{v}_{ij} - (\vec{v}_{ij}\cdot \vec{e}_{ij})\,\vec{e}_{ij}|.
\end{equation}
We further assume
that the fluid mixing on scales below the particle cells can be
approximately described as a diffusion process, operating with
diffusion constant $D= \chi L^2 / t_{\rm KH}$, where $\chi$ is a
dimensional efficiency that controls the strength of the mixing (and
which needs to be determined empirically).  We hence effectively model
the mixing with heat diffusion of the form $\partial u /\partial t = D
\nabla^2 u$.

Using the SPH discretization of thermal conduction as a guide
\citep{Jubelgas2004}, we can readily find a discretization of the heat
diffusion for the Voronoi particle discretization. We obtain
\begin{equation}
m_i \frac{{\rm d}u_i}{{\rm d}t}
 = 2\pi\chi \sum_j A_{ij} (1- \bar{f}_{ij}) |v_{ij}^{\parallel}| \sqrt{\rho_i\rho_j} (u_j -
 u_i).
\label{EqnDiff}
\end{equation}
Here we introduced a further factor $(1- \bar{f}_{ij})$ in a similar
manner as in (\ref{BalsaraF}). This Balsara-like factor is used to
restrict the diffusion only to areas where the compression (as
measured by $|\vec{\nabla} \cdot \vec{v}| $) is clearly negligible
compared to the shear (as measured by $|\vec{\nabla} \times \vec{v}|
$).

Note that this equation preserves the total thermal energy, and heat
energy only flows from hotter to colder particles. The corresponding
rates of entropy change for each particle can be obtained by
multiplying with $(\gamma-1)/ \rho_i^{\gamma-1}$. While the specific
entropy of
individual particles may go down if they give up some of their heat
energy, the total entropy of the system increases due to this process,
which can be interpreted as providing the necessary mixing entropy.

One important difference of our parameterization of `artificial heat
conduction' to the model of \citet{Price2008} is that the mixing only
occurs in shear flows, and that contact discontinuities without shear
are hence not affected.

We note that due to the parabolic character of the diffusion problem,
it can be problematic to integrate equation (\ref{EqnDiff}) with an
explicit time integration scheme, since the von Neumann criterion
imposes relatively small timestep limits.
\begin{equation}
\Delta t \, \leq \, \frac{\left( \Delta x \right)^2}{D} \, \approx \,
\frac{\left( \Delta x \right)}{ 2 \pi \chi |v_{ij}^{\parallel}|},
\end{equation}
where we estimated $\min\left( \frac{\rho_i + \rho_j}{2\,   \sqrt{\rho_i  \rho_j}}\right) = 1$
 and assumed $\Delta x \approx L$. The CFL criterion is more
 restrictive than this timestep as long as
\begin{equation}
 |v_{ij}^{\parallel}| \,  \ll \,  \frac{1}{2 \pi  \chi}   \, c_{\mathrm{sound}},
\end{equation}
which is usually the case and hence not too restrictive. Indeed, so
far we have not encountered problems with the explicit time
integration scheme that is implemented for the mixing at present. If
needed, an implicit scheme with perfect stability could however be
easily adopted, like the one discussed in \citet{Petkova2009}.

\section{Issues of Cell Regularity}  \label{SecRegularity}

A common feature of particle hydrodynamic schemes is their ability to
automatically provide an adaptive resolution. As a result, dense regions
are modelled with better accuracy thanks to their smaller mean distance of
particles.  But besides the particle number density the regularity
of the Voronoi cells is an important factor in determining the achieved
precision, as shown in Appendix~\ref{Test of accuracy}, where we give
quantitative results for the accuracy of our gradient estimates as a
function of the shape distortions of cells.  Highly irregular,
sliver-like Voronoi cells may also lead to very small, computationally
costly timesteps, because the permissible timestep size is effectively
proportional to the distance to the nearest neighbour.

Connected to this problem is the issue of how to safely prevent
inter-particle penetrations, which is required for a proper
representation of the fluid with its single valued velocity field. If
two particle approach each other rapidly, it is possible that the
particles pass through each other unless this is prevented with a
sufficiently strong artificial viscosity. If the Voronoi mesh is very
irregular and features a large number of close particle pairs, it
becomes more difficult to ensure this, simply because rather large
viscous forces that act over short timescales are required to prevent
the small particle separations from becoming still smaller.  These
problems are significantly alleviated if the tessellation is relatively
`regular', i.e.~if cells have a small aspect ratio, and if their
generating points lie close to the centroids of the corresponding cells.

\begin{figure}
\begin{center}
  \includegraphics[width=0.5\textwidth]{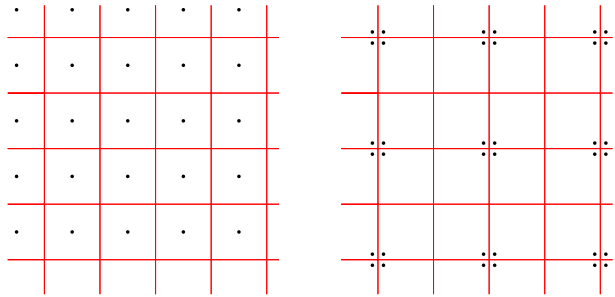}
\caption{Two point distributions and their corresponding Voronoi
  tessellations. The important point illustrated by this example is that
  {\em different} point distributions may have {\em identical} Voronoi
  tessellations.}
\label{FigDegeneracy}
\end{center}
\end{figure}

Figure~\ref{FigDegeneracy} illustrates another important feature of
Voronoi meshes, which we may perhaps call `mesh degeneracy'.  In this
example, the mesh for two different point distributions is shown, but in
both cases an identical Cartesian Voronoi mesh results, hence the
density and pressure estimates are both equal. In fact, one can continue
to move the groups of four points around the mesh vertices in a mirrored
fashion arbitrarily close towards the corners of the mesh, without
changing the situation. If one now imagines adding some random velocity
field to the points when they are very close, it is clear that it will
be much harder to prevent an erroneous particle crossing in the case
where the points are far from the centres of their associated cells than
in the case where they sit right at these centres. 

One might argue that situations as shown in Figure~\ref{FigDegeneracy}
are artificial and hence do not affect the simulation of flows where
Voronoi diagrams are seldom that regular. But situations occur where 
the volume is only slightly increased when particles approach. Also
considering a finite timestep these particles' resistance against a
clumping or even interpenetration is then uncomfortably weak.

If possible, it would therefore be desirable to formulate the dynamics
such that the mesh automatically maintains a certain degree of
regularity. We note that this cannot be expected to happen by itself for
the density estimation scheme we implemented thus far. This is again
made clear by the example in Figure~\ref{FigDegeneracy}, where the
pressure gradient vanishes in both cases if the specific entropies and
masses of all particles are the same. Below we therefore consider two
possible approaches to introduce small correction forces into the
dynamics with the goal to rearrange the points to achieve a more regular
tessellation.

\subsection{Viscous forces that help to improve order} \label{PPO}

Experience with our new VPH scheme shows that especially in highly
irregular, turbulent flows and in situations with strong gravitational
forces some cells can become quite irregular. In this context, we loosely
define an irregular cell as one whose generating point is substantially
displaced from the centroid of the cell and/or whose aspect ratio is
quite high, i.e.~a cell with a comparatively large ratio of surface area
to volume. In this subsection we consider a scheme where the artificial
viscosity is modified such that it serves a second purpose, namely to
have the tendency to make cells `rounder'.

To this end we introduce the notion of a `partial pressure' for each of
the pyramids that make up a Voronoi cell.  These pyramids are spanned by
the Voronoi faces and the defining point of the cell, which acts as
their apex.  We define the `partial pressure' of the pyramid of cell $i$
facing cell $j$ in terms of its volume $V_{ij} = A_{ij} R_{ij}/6$ (or
$V_{ij} = A_{ij} R_{ij}/4$ in 2D) and by assigning a share $m_{ij} = m_i
{A_{ij}} / {\sum_{k} A_{ik}}$ of the cell's mass to the pyramid,
i.e.~its mass fraction is taken to be proportional to its contribution
to the total surface area of the cell. This yields
\begin{equation}
  P_{ij} = s_i \rho_{ij}^\gamma =s_i \left( \frac{6\,m_i}{ R_{ij} \:
    \sum_{k} A_{ik}} \right)^\gamma.
\end{equation}
The idea is now to define an additional viscous force between a pair of
particles arising from the difference of this partial
pressure to the full pressure of the cell. This will lead to
rearrangements of the points until the differences in the partial
pressures of each pyramid to that of the cell become small, which
happens when the point is approximately equidistant to all the faces of
its cell, implying a regular cell shape.

We hence make the ansatz
\begin{equation}
(\vec{f}_{\rm order})_{ij} = -\kappa A_{ij} (P_{ij} - P_i + P_{ji} - P_j) 
\frac{\vec{e}_{ij}}{2} 
\end{equation}
for `ordering forces' between a pair of particles, 
with the total force on particle $i$ being given by
\begin{equation}
 m_i (\vec{a}_{\rm order})_i =  \sum_{j\ne i} (\vec{f}_{\rm order})_{ij} .
\end{equation}
Here $\kappa$ is a dimensionless parameter describing the strength of
the effect. These forces are antisymmetric, and in general can be both
of repulsive and attractive nature. In order to maintain total energy
conservation, the work of these forces needs to be balanced in the
evolution of the entropies of the particles, similar to what is done
for the ordinary artificial viscosity. This yields an additional
contribution to the entropy change of the form
\begin{equation}
\frac{\mathrm{d} s_i}{\mathrm{d} t} =
\frac{(\gamma -1)}{ \rho_i^{\gamma-1}} \frac{1}{2} \sum_j (\vec{f}_{\rm
  order})_{ij} \cdot \vec{v}_{ij} .
\end{equation}
Note that a small local decrease of the entropy sum can result in
principle if order is restored in the particle distribution, but this
effect is small and has played no role in all our tests. In refinements
of the above scheme, it is also possible to make $\kappa$ spatially and
temporarily variable. For example, we have typically used this scheme in
a combination with a switch that only sets $\kappa > 0$ if there is a
strong local compression, because that is where cell shapes distort the
most. 

When discussing results, we will refer to this method as `partial
pressure ordering' or PPO, whereas the method for improved cell
regularity discussed in the next subsection will be referred to as
`shape correction forces'. If none of these additional schemes to
regularize cell shapes is employed, we simply refer to the method as the 
`plain Voronoi scheme'.

\subsection{Imposing regularity through the fluid Lagrangian} \label{SecShapeCorrect}

If irregular cell shapes occur, we ideally would like that small
adjustment forces appear naturally that tend to make the mesh more
regular again. These adjustment forces should preserve the energy and
momentum conservation of the scheme. This will automatically be the case
if they are derived from a suitably defined Lagrangian or Hamiltonian.
We are hence led to modify the fluid Lagrangian slightly to include
factors that penalize highly distorted cell shapes. The idea is that
such distorted cells should raise the estimate of the inner energy slightly,
such that they become energetically disfavored.

We consider two ways to measure shape distortions of cells, which may
either be used individually, or combined.  One is based on the
displacement of a point from the centroid of its associated cell. The
idea here is that it is advantageous if a point stays close to the
center of a cell. In particular, as we will discuss in more detail in
Section~\ref{Dispersion-relations}, it turns out that the ordinary VPH
scheme is not able to support waves in regular Cartesian grids at the
Nyquist frequency, a deficit that could be cured if there is always a
(weak) restoring force if a point is displaced from the centre of its
Voronoi cell.

The other is to measure the shape directly, and to steer the particle
motion such that high aspect ratios are avoided. We construct a shape
measure based on the second moment of the cell, which we compare to a
suitably defined cell radius. This measure will have a minimum for
`round cells', while severe distortions from roundness (like highly
elongated cells) should trigger restoring forces.

Both of the above measures of cell regularity can be introduced into the
fluid Lagrangian by multiplying the thermal energy with correction factors
that increase the energy slightly if a point is displaced from the
centroid, or if a cell is elongated. Specifically, we adopt as
Lagrangian
\begin{eqnarray}
L & = & \sum_k \frac{1}{2} m_k\dot\vec{r}_k^2 \quad - \label{eqnLgr} \\
&& \hspace*{-0.5cm}\sum_k
\frac{P_kV_k}{\gamma-1}\left[1+\beta_0\frac{(\vec{r}_k-\vec{s}_k)^2}{V_k^{2/d}}\right]
\left\{    1+\beta_1 \left( \frac{\vec{w}_k^2}{V_k^{2/d}} -
\beta_2\right)    \right\}. \nonumber 
\end{eqnarray}
Here $\vec{r}_k$ is the coordinate of a point, $V_k$ is the volume of
its corresponding Voronoi cell, and $\vec{s}_k$ is the cell's
centroid. $P_k$ is the ordinary pressure of the cell, where we set
$\rho_k = m_k/V_k$ as usual. The coefficients $\beta_0$ and $\beta_1$
are introduced to measure the strength of the correction forces
associated with offsets from cell centres, or with high aspect ratios,
respectively. For $\beta_0 = \beta_1 =0$, the ordinary fluid Lagrangian
of the VPH scheme is recovered.

We define the centroid of a cell as
\begin{equation}
\vec{s}_k \equiv \left<\vec{r}\right>_k = \frac{1}{V_k}\int 
\vec{r}\, \chi_k(\vec{r})\, {\rm d}\vec{r} ,
\end{equation}
where $\chi_k$ is the characteristic function of cell $k$,
i.e.~$\chi_k(\vec{r})=1$ if the point $\vec{r}$ lies in the cell $k$,
and $\chi_k(\vec{r})=0$ otherwise.  The shape of a cell is measured via
the second moment
\begin{equation}
\vec{w}_k^2 \equiv \left<(\vec{r}-\vec{s}_k)^2\right>_k = \frac{1}{V_k}\int
    (\vec{r}-\vec{s}_k)^2
\chi_k(\vec{r}) \, {\rm d}\vec{r}.
\end{equation}
The factors in squared and curly brackets in the Lagrangian of equation
(\ref{eqnLgr}) are the adopted energy correction factors that raise the
energy of distorted cells. Here $d$ counts the number of dimensions,
i.e.~$d=2$ for 2D and $d=3$ for 3D. The factor $V_k^{2/d}$ is hence
proportional to the `radius' $R_k = V_k^{1/d}$ of a cell squared.
$\beta_0$ measures the strength of the effect of displacements of points
from the centroid of a cell, while $\beta_1$ is the corresponding factor
for the aspect-ratio factor.  The constant $\beta_2$ is only introduced
to prevent that even round cells lead to a significant enhancement of
the thermal energy. For perfectly round cells, we expect in 2D
approximately circles for which $w_k^2 = V^{2/d} / (2\pi)$, hence we
pick $\beta_2 = 1/(2\pi)$. In 3D, we have spherical shapes instead and
we pick $\beta_2 = 3/5 (3/4\pi)^{2/3}$.

The equations of motion for the Lagrangian (\ref{eqnLgr}) can be
derived in closed form for the Voronoi mesh, but due to the length of
the resulting expressions we give their derivation in
Appendix~\ref{AppShape}. The advantage of using the Lagrangian to
obtain the cell-shaping forces is that the scheme then still
accurately conserves total energy, momentum and entropy, while at the
same time remaining translational and rotationally invariant. Also,
the correction forces are `just right' to achieve the desired
regularity of the mesh, something that is difficult to achieve with
any heuristic scheme to derive such forces, like the one we tried in
the previous subsection.

In our results section, we show that this method is indeed capable of
maintaining nicely regular meshes in the sense described above. However,
we also caution that the stronger the extra forces are, the more unwanted
features start to appear as well. First of all, the extra forces may
introduce subtle deviations from the dispersion relation of an ideal
gas, and may lead to spurious motions in situations with pressure
equilibrium. The second, probably more serious side effect of this
method may occur when the cells cannot easily relax to the desired
regular cell structure, for example along a strong jump in density.  In
this case a pressure anomaly may develop due to the cell-shaping forces,
similar to what is found in SPH across contact discontinuities. Still,
we find that moderate values of $\beta_0$ and $\beta_1$ help to improve
the accuracy of the VPH scheme without distorting the inviscid dynamics
of an ideal gas too much.

\section{Implementation} \label{SecImplementation}

We have implemented the above hydrodynamical particle model into the
cosmological TreeSPH simulation code {\small GADGET-3}, an updated
version of {\small GADGET-2} \citep{gadget2,gadget1}. This code is
parallelized for distributed memory machines, and offers
high-performance solvers for self-gravity as well as individual and
adaptive timestepping for all particles. Our strategy in our
modifications has been to implement Voronoi-based particle hydrodynamics
as an alternative to SPH within the {\small GADGET-3} code. This is, in
particular, ideal for facilitating comparisons between SPH and our
Voronoi-based scheme, and it also allows us to readily use all the
non-standard physics already implemented in {\small GADGET-3}
(e.g.~radiative cooling and star formation) for calculations with
Voronoi-based fluid particle dynamics.

The primary new code needed in {\small GADGET-3} is an efficient mesh
construction algorithm. To this end we adapted and modified the parallel
Delaunay triangulation engine from the {\small AREPO} code
\citep{AREPO}, and turned it into an optional module of the SPH code
{\small GADGET-3}. In brief, the tessellation code uses an
incremental construction algorithm for creating the Delaunay
tessellation. Particles are inserted in turn into an already
existing, valid tessellation. To this end, in a first step the tetrahedron
in which the new point falls is located, and then it is split into several new
tetrahedra, such that the inserted point becomes part of the
tetrahedralization. However, some of the new tetrahedra may then not
fulfill the empty-circumsphere property, i.e.~the tessellation is not a
Delaunay triangulation any more. Delaunayhood is restored in a second
step by local flip operations that replace two adjacent tetrahedra 
with three tetrahedra, or vice versa, until all tetrahedra fulfill 
again the empty-circumsphere property. At this point, the next particle 
can be inserted. We have implemented the mesh construction both for 3D
and 2D within the {\small GADGET-3} code and parallelized it for
distributed memory machines.

At the beginning, a large tetrahedron is constructed that encloses the
full computational domain. The boundary conditions (always adopted as
periodic at the moment) of the rectangular computational domain as well
as the boundaries arising from the domain decomposition are treated with
`ghost particles'.  One technically difficult aspect is to make the
tessellation code completely robust even in the case of the existence of
degenerate particle distributions, where more than 4 points lie on a
common circumsphere (or more than 3 points are on a common
circumcircle). Detecting such a case robustly and correctly in light of
the finite precision of floating point arithmetic is a non-trivial
problem. However, the incremental insertion algorithm requires
consistent and correct evaluations of all geometric predicates,
otherwise it will typically fail in situations with degeneracies or
near-degeneracies. We solve this problem by monitoring the floating
point round-off in geometric tests, and by resorting to exact arithmetic
in case there is a risk that the result of a geometric test may be
modified by round-off error.

\section{Test Results} \label{SecResults}

In this section, we discuss a number of test problems carried out with
 our new hydrodynamical particle method, focusing in particular on regimes
where differences with respect to SPH can be expected. An application
of the method in full cosmological simulations of galaxy formation
will be presented in future work.

\subsection{Surface tension} \label{SecSurfTens}

In standard entropy-conserving SPH with particles, there is a subtle
surface tension effect across contact discontinuities with a large jump
in density. This can be understood as a result of the desire of SPH to
suppress mixing of the two phases, because this is energetically
unfavorable for fixed particle entropies. For the mixed state,
approximately the same average density would be estimated for each
particle, which leads to a higher estimate of the thermal energy, unless
the thermodynamic entropies are averaged between the particles as well,
which is an irreversible process in which entropy is in fact produced if
the total energy stays constant.

\begin{figure*}
\begin{center}
  \includegraphics[width=0.8\textwidth]{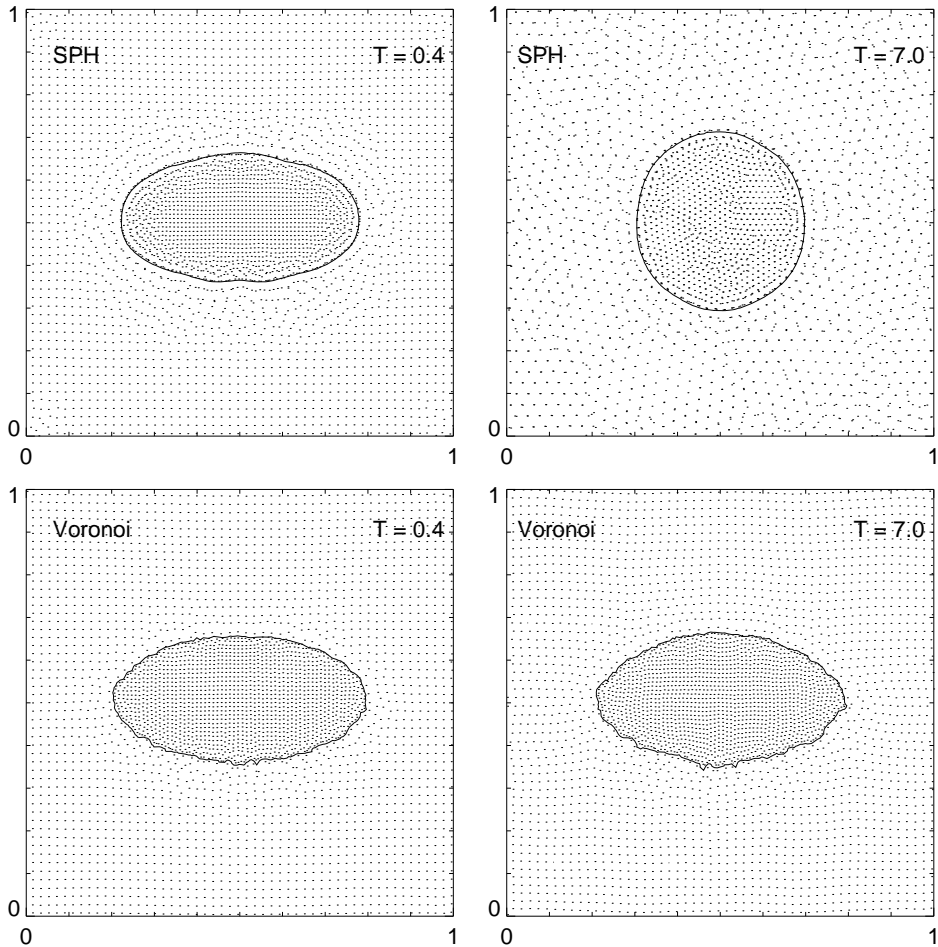}
\caption{Surface tension effect in SPH. The left column shows an overdense
  ellipsoidal region shortly after it is set-up at $t=0$ in pressure equilibrium
  within a thinner background. When evolved with SPH, the ellipsoid
  slowly transforms into a sphere, as shown by the state of the system
  after time $t=7$ (top left). In contrast, the Voronoi scheme can
  preserve the shape much better (bottom panels) and shows no sign of
  surface tension effects.}
\label{FigEllipse}
\end{center}
\end{figure*}

To demonstrate the existence of the surface tension effect, we have
prepared (in 2D for simplicity) an overdense ellipse in a thin
background medium, at pressure equilibrium. For definiteness, the
density of the ellipse was set to $\rho_2=4$, that of the background
medium to $\rho_1=1$, with a pressure of $P=2.5$. 3854 equal mass
particles in a periodic box of unit length on a side were used to set
up the experiment. The particles have been arranged on a coarse
Cartesian grid in which an ellipsoidal region was excised. This region
has then be filled with a finer Cartesian grid. The specific entropies
of the two sets of particles were initialized differently such that an
equal pressure for the two phases results. No attempt was made to
somehow soften the transition between the two
phases. Figure~\ref{FigEllipse} shows the initial configuration, as
well as the particle distribution after a time $t=7$, both for SPH and
for the Voronoi-based fluid particle approach.

Even though the pressures of the particles are formally equal for all
particles in the initial conditions, the ellipse transforms to a circle
when SPH is used.  In contrast, for the VPH scheme, the same experiment
maintains the initial shape of the ellipse, modulo some small
rearrangements of the points near the boundary, since the initial set-up
was not in perfect equilibrium (due to the fact that the point
distributions of the two Cartesian grids used to set up the two phases
do not match seamlessly at the boundary). Clearly, the numerical
realization of the contact discontinuity in SPH gives rise to a spurious
surface tension, and this in turn will suppress Kelvin-Helmholtz
instabilities below a certain critical wavelength. The VPH approach does
not have this problem and can in principle accurately support a contact
discontinuity at each face boundary between individual cells. It needs
to be stressed however that also in the Voronoi scheme no mixing of the
entropies at the particle level happens. If the particles of two phases
were simply spatially mixed while keeping their specific entropies
constant, the resulting medium would not be at a single temperature or
density. {\bf We note that the effect is also present for a set-up with
  unequal particle masses. In this case, the variationally derived
  entropy-conserving version of SPH will however lead to a change of the
  effective number of neighbours across the contact discontinuity,
  because it keeps the mass in the kernel volume constant.}

\subsection{Sod shock tube}

The classic Sod shock tube tests examine the ability of a hydrodynamic
scheme to reproduce the basic wave structure that appears in the Riemann
problem, namely shock waves, contact discontinuities and rarefaction
waves. Also, comparison to the analytic solution gives a useful
quantitative benchmark for the accuracy of a scheme.

We consider gas that is initially at rest. In the left half-space, the
pressure is $P_1 = 1.0$ and the density is $\rho_1 = 1.0$, whereas in
the right half-space we adopt $P_2 = 0.1795$ and $\rho_2 = 0.25$. The
adiabatic index is set to $\gamma = 1.4$. The same sod shock
parameters have previously been used in a number of code tests
\citep[e.g.][]{Hernquist1989,Rasio1991,Wadsley2004,gadget2}.
When the evolution begins, a shock wave of Mach number $M=1.48$
travels into the low-pressure region, and a rarefaction fan moves into
the high pressure region. In between, a moving contact discontinuity
develops.

\begin{figure}
\begin{center}
\vspace{-9mm}

\hspace{-7mm}
 
\includegraphics[width=0.51\textwidth]{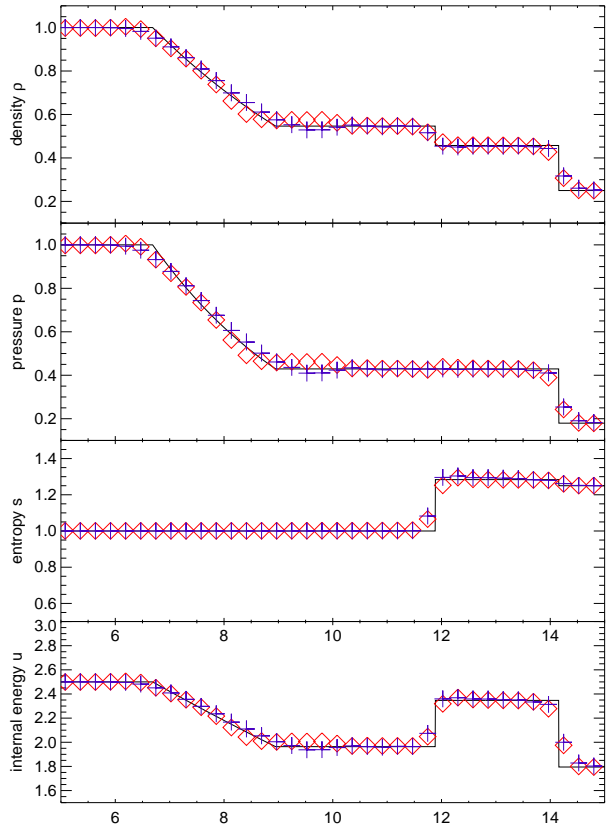}
\vspace{-5mm}
\caption{3D Sod shock tube simulation at $t=3.0$. We compare results
  from our Voronoi particle scheme (blue crosses) and the SPH result 
  with 32 neighbours (red diamonds) with the analytical
  solution (solid lines) in terms of the density, pressure, entropy
  profiles and internal energy.  \label{sod_shock} }
\end{center}
\end{figure}

In our numerical test of this problem with the VPH scheme we use a 3D
setup in a box with dimensions $(20,1,1)$, where the left and right
halves are filled with particles arranged on a Cartesian grid.
Altogether 8370 particles with equal masses were used as initial
condition. The evolution was then carried out with our default
settings for the artificial viscosity until $t=3$.  In
Fig.~\ref{sod_shock}, we compare the numerical result to the
analytical solution at this time. Reassuringly, we find quite good
agreement of the VPH scheme with the analytical solution. In
comparison to SPH, the rarefaction wave in the Voronoi simulation
shows a slight dip at the low density end, but not the hump at the
high density end that is typical of SPH. Another difference is that
the contact discontinuity is sharper and better preserved in the VPH
approach.  All of the described features appear to be largely
independent of how the points are distributed initially as long as the
gas is relaxed on both sides. In particular, the arrangement on a
Cartesian grid does not lead to any noticeable artifacts compared with
simulations where the initial point distribution is less ordered and
has a glass-like configuration. However the results are somewhat less
accurate for irregular grids (see also \ref{Test of accuracy}) when
the plain Voronoi scheme is used.

We have also examined the influence of the extra forces that can be
enabled to improve the regularity of the mesh (see Sections~\ref{PPO}
and \ref{SecShapeCorrect}). To this end we tested the effect of the
PPO scheme as well as the shape correction method with parameters up
to $\beta_{0} = 1.2, \, \beta_{1} = 0.1$.  When these ordering forces
are invoked, the results for the sod-shock test are in general not
influenced much, but the particle noise around the analytical solution
is reduced.  We also found that the extent of postshock oscillations
for weaker artificial viscosity settings tends to be reduced if these
ordering methods are used.

\subsection{Dispersion relations} \label{Dispersion-relations}

Even though this may seem like a simple test, it is actually important
to check how well our new method can simulate
small-amplitude\footnote{The amplitude needs to be small in order to
  prevent wave steepening.} acoustic waves, especially at low resolution
when few points per wavelength are available. We are especially
interested in how accurately the expected dispersion relation is
reproduced in this regime, i.e.~whether such waves propagate with the
correct speed of sound. A secondary question is how strongly such waves
are damped by the artificial viscosity in the scheme.

\begin{figure}
\begin{center}
\vspace{-5mm}
\hspace{-11mm}
  \includegraphics[width=0.53\textwidth]{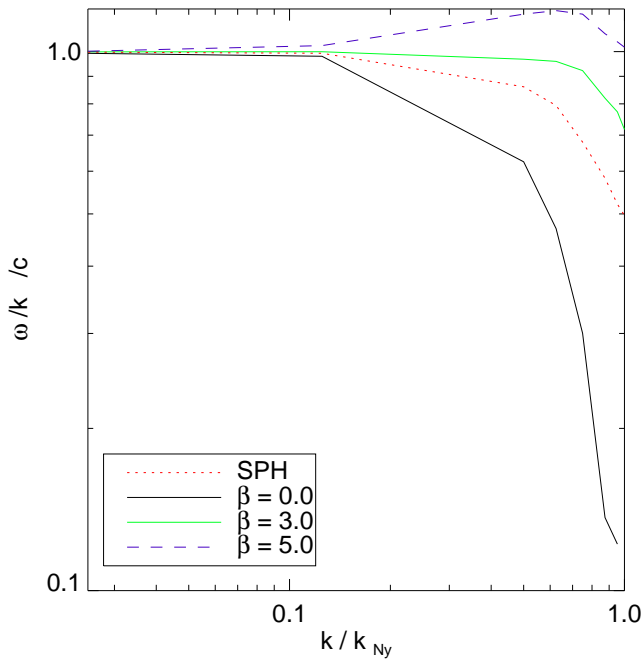}
\vspace{-5mm}
\caption{Dependence of the numerical sound speed on wavenumber,
  expressed in units of the Nyquist frequency of the underlying particle
  grid. For the standard VPH scheme, waves at the Nyquist frequency are
  not propagated properly (hence no frequency can be measured at
  $k=k_{\mathrm{Ny}}$), but this is remedied by the shape correction
  forces. If they are invoked, the resulting dispersion relation becomes more
  accurate than that of SPH for all~$k$. \label{disp_rel}}
\end{center}
\end{figure}

To measure the dispersion relation, we set up small-amplitude standing
waves in a periodic box and measure their oscillation frequency. In
Figure~\ref{disp_rel}, we compare results for SPH with our Voronoi-based
fluid particle model, with and without shape correction forces, as a
function of wavenumber. The wavenumber is normalized to the Nyquist
frequency of the initial particle grid, such that $k/k_{\rm Nyquist}=1$
corresponds to the shortest wave that can be represented by the
particles. In this standing wave, neighbouring particles oscillate 180
degrees out of phase `opposite' to each other.

A first important result made clear by Figure~\ref{disp_rel} is that for
the standard VPH scheme the oscillation frequency for $k/k_{\rm
  Nyquist}=1$ drops to zero, or in other words, such waves are not
supported by the scheme at all. This is readily understood from the
degeneracy effect pointed out in Figure~\ref{FigDegeneracy}. If
particles are set-up such that they `collide' in a pairwise fashion,
then there is nothing in the reversible part of the dynamics of the VPH
scheme that can prevent an interparticle penetration, simply because the
pressure gradient stays zero in this case.  However, this situation is
exactly the one encountered if we prepare a standing wave at the Nyquist
frequency of an initially regular particle grid. The wave will not
oscillate since the pressure gradient will remain zero, and therefore
particle crossings would be inevitable (unless prevented by the
artificial viscosity). This is potentially a serious shortcoming of the
VPH scheme in its standard form, as it means that it cannot treat waves
at around the Nyquist frequency properly.

However, the shape correction force due to $\beta_0$ has exactly the
right property to make these small waves oscillate again. In fact, we
can calculate what value of $\beta_0$ is required to reproduce the
dispersion relation at $k/k_{\rm Nyquist}=1$ exactly. For this value
of $\beta_0\simeq 5.5$, we however also get slightly too stiff
behaviour of the fluid for somewhat longer wavelengths, as shown by
Fig.~\ref{disp_rel}. A value of around $\beta_0\sim 3$ represents a
good compromise, and in particular yields a more accurate dispersion
relation than SPH for all $k$. We also note that for certain numbers
of neighbours, the SPH result is inaccurate at {\em all} wavelength;
here the numerical soundspeed shows an offset relative to the expected
sound speed, which is presumably a result of a bias in the density
estimate for the background density.

\subsection{Density noise and regularity in a settled particle distribution}

In this subsection we want to examine the level of noise present in a
relaxed region of gas of constant specific entropy, as it may arise
somewhere within a larger, self-consistent simulation.  To mimic this
situation, we start from a distribution of points arranged on a
Cartesian grid and impose a random Gaussian velocity field with
dispersion $\langle v^2 \rangle= 0.05\,c_s^2$ and zero mean. The idea
is that these velocity fluctuations break the initial grid symmetry
and will then get damped away by the artificial viscosity, which is
here set to a high value to speed up the process of settling to a new
pressure equilibrium. Since we want to retain the initially equal
values of the specific entropies per particle, we disable the entropy
source term for the viscosity in this experiment. Once the new
equilibrium for an irregular particle distribution is achieved, we
can then examine the noise properties of this particle representation
of a constant density, constant pressure gas.

For SPH with $N=16$ neighbours, we find that the particles settle into
several domains in which the points are quite regularly distributed,
based on visual inspection. The estimated density values $\rho_i$ for
the particles are {\em not} all equal though, instead they show a
distribution with rms-scatter equal to $\sim 1.4\, \%$, and also a
small {\em bias} relative to the expected value equal to the mean
density $\left<\rho \right> = Nm/V$ of the full volume.

\begin{figure}
\begin{center}
\hspace*{-22mm}
  \includegraphics[width=0.5\textwidth]{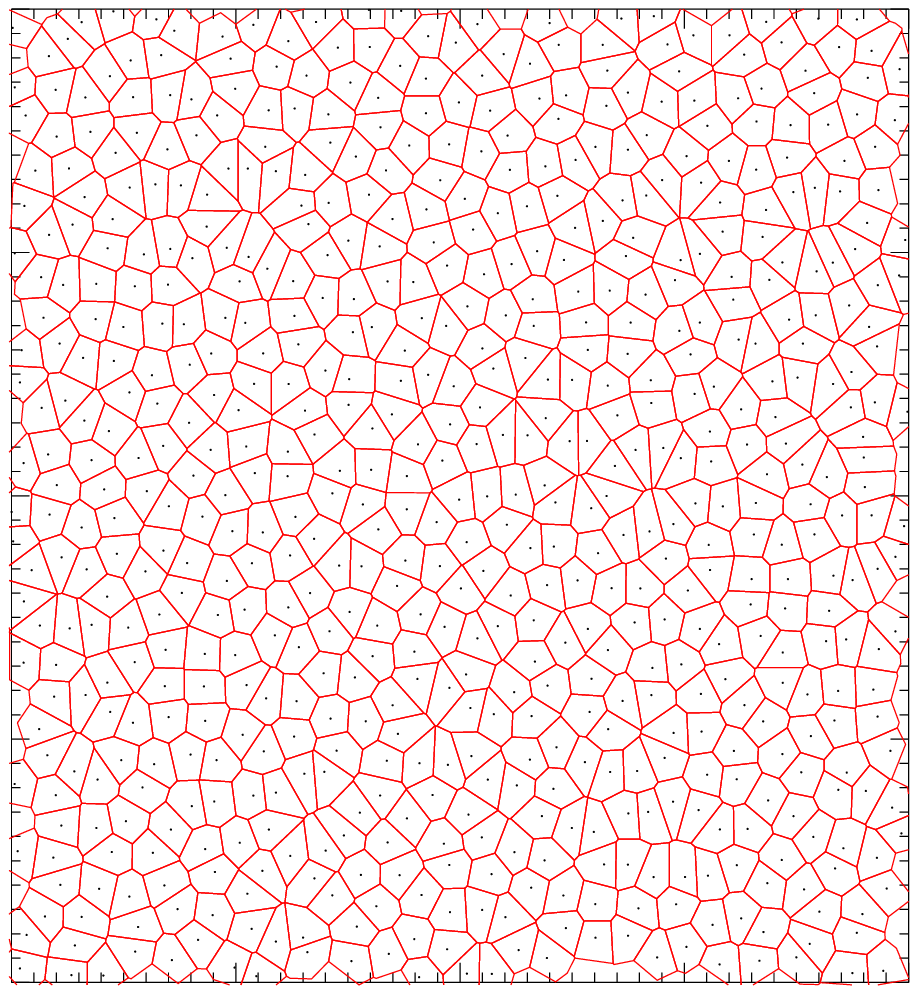}\\
	\vspace*{-12mm}
\hspace*{-22mm}
\includegraphics[width=0.5\textwidth]{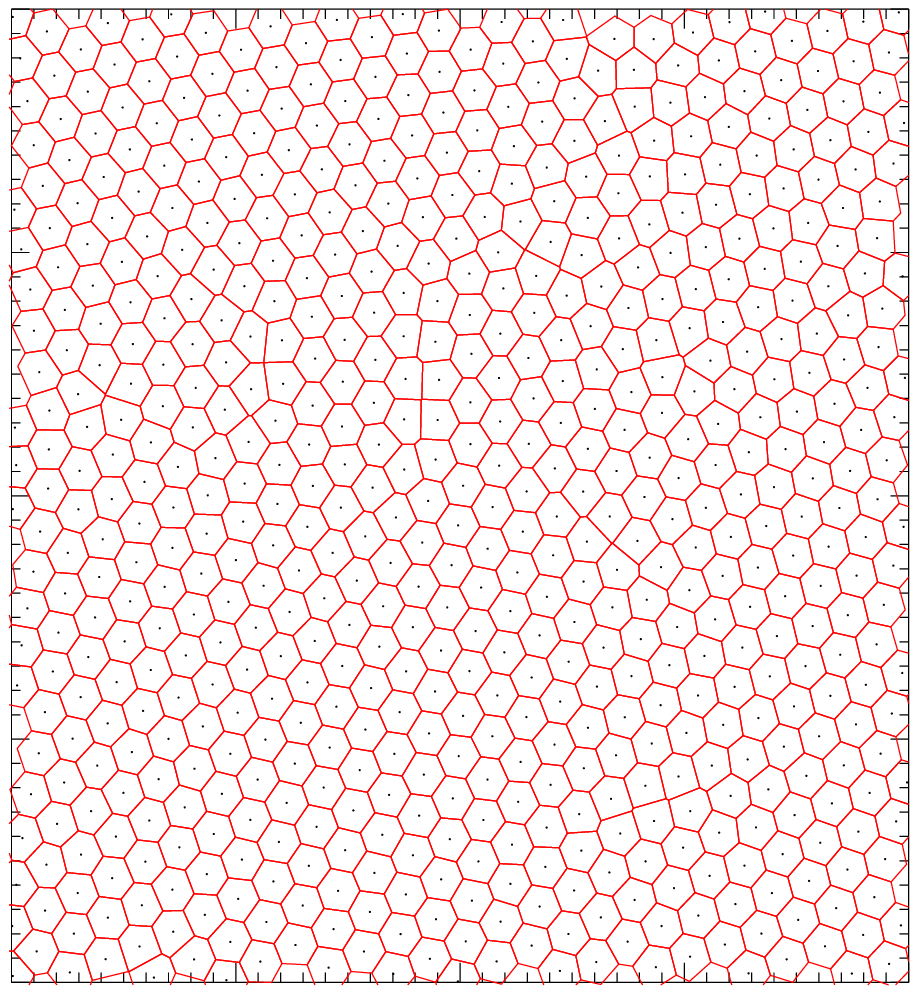}\\
\vspace*{-8mm}
\label{tessellation_plot}
\caption{Final mesh geometry in VPH in a 2D settling test, carried
  out without (top) or with shape correction forces (bottom) based on
  $\beta_0=1.0$ and $\beta_1=0.01$.}
\label{FigSettling}
\end{center}
\end{figure}

In contrast, the standard VPH approach creates a distribution in which
the density values are essentially single-valued, and are all very close to
$\left<\rho \right>$. This means that the cells have all equal volume,
and the residual pressure fluctuations, if any, are extremely
small. However, the geometry of the Voronoi tessellation is quite
irregular and features numerous cells with relatively large aspect
ratios, or with points close to cell boundaries. This can be seen in
Figure~\ref{FigSettling}, where we show a plot of the final mesh for a
test case carried out in 2D.

\begin{figure}
\begin{center}
  \includegraphics[width=0.45\textwidth]{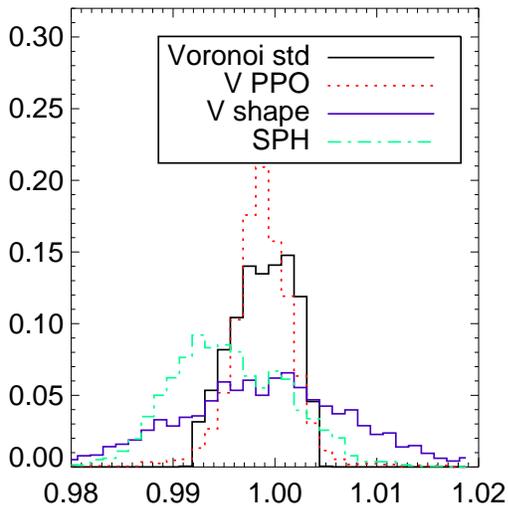}
\caption{Density distribution functions of the particles in a settling
  test for constant entropy gas, carried out with different schemes. We
  compare SPH (red), ordinary VPH (black), the PPO version of VPH
  (blue), and VPH with shape correction forces based on $\beta_0=1.0$
  and $\beta_1=0.01$ (in green). The distributions were measured at a
  time when the initial kinetic energy had decayed to $E_{\mathrm{kin}}
  \approx 0.001 E_{\mathrm{kin}}(t=0)$.  
\label{FigSettling2}}
\end{center}
\end{figure}

It is now interesting to repeat the test for the case when shape
correction forces according to Section~\ref{SecShapeCorrect} are
included. As desired, the final mesh becomes much more regular in this
case, as seen in the corresponding example included in
Figure~\ref{FigSettling}.  However, even in the final equilibrium
state the correction forces do not necessarily completely vanish in
this case. Instead, they are compensated by small residual pressure
(and hence also density) fluctuations. This is demonstrated in the
distribution functions of the density values for the three cases we
considered, which we give in Figure~\ref{FigSettling2}. Here the shape
correction case yields a somewhat broader distribution, similar to
SPH, but without showing a bias.

\begin{figure}
\begin{center}
  \includegraphics[width=0.45\textwidth]{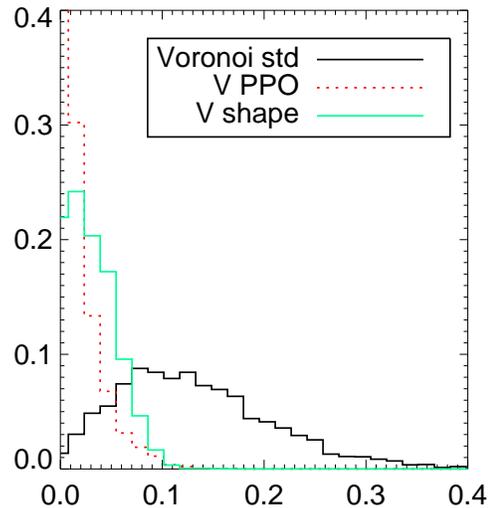}
\vspace*{-8mm}
\caption{Cell-regularity of a noisy flow after relaxation.  To
  characterize the regularity of the cells, we simply consider the
  distribution of the distance of the points to their cell's
  centroids, in units of the 2D cell radius $r=\sqrt{{V}/{\pi}}$.
  The black histogram shows the distribution for the ordinary Voronoi
  scheme, blue shows the PPO version, and green lines give the result
  for the Voronoi with shape correction forces derived from the
  Lagrangian.  The distributions were measured at a time when the
  initial kinetic energy had decayed to $E_{\mathrm{kin}} \approx
  0.001 E_{\mathrm{kin}}(t=0)$.  }
\label{FigSettling3}
\end{center}
\end{figure}

In Figure~\ref{FigSettling3}, we give a quantitative measure for the
cell-regularity (here taken as the distribution function of the
normalized displacement of points from the centres of their cells) of
the final meshes in the Voronoi-based simulations.  Even moderate
values of the coefficients $\beta_0$ and $\beta_1$ can drastically
improve the regularity of the particle distribution while introducing
less noise in the density estimate than anyway present in SPH.

\subsection{Point explosion}

If energy is injected at a point into a cold gas at constant density, a
spherical blast wave will develop. The Taylor-Sedov solution provides an
analytic solution for this self-similar problem, which is a useful test
involving very strong shocks. We have set-up this {\bf problem in 3D},
using unit background density (represented with a Cartesian mesh),
$\gamma=5/3$ and vanishingly small initial specific entropy compared to
the injected energy of $E=1$.  We inject the energy into the centre of
the domain at time $t=0$.  To avoid that the evolution is strongly
affected by the non-spherical geometry of the central Voronoi cell, we
have spread out the energy with a Gaussian kernel with a radius of about
4 mean particle spacings.

We note that this set-up can be especially sensitive to the problem of
particle crossing when a too low viscosity and individual timesteps are
used. In the latter case, the local Courant timestep of particles
outside of the explosion is initially very big. When the supersonic
shock front arrives, such particles may then still live on a too large
timestep, such that they are effectively overtaken by the shock,
creating severe artifacts in the evolution of the shock front. We have
addressed this in our test by imposing a low enough maximum timestep
for all particles, but more sophisticated schemes to set the
timesteps, which guarantee that it is reduced before the shock
arrives, can of course be implemented in principal
\citep[see][]{Saitoh2008,AREPO}.

\begin{figure}
\begin{center}
\includegraphics[width=0.4\textwidth]{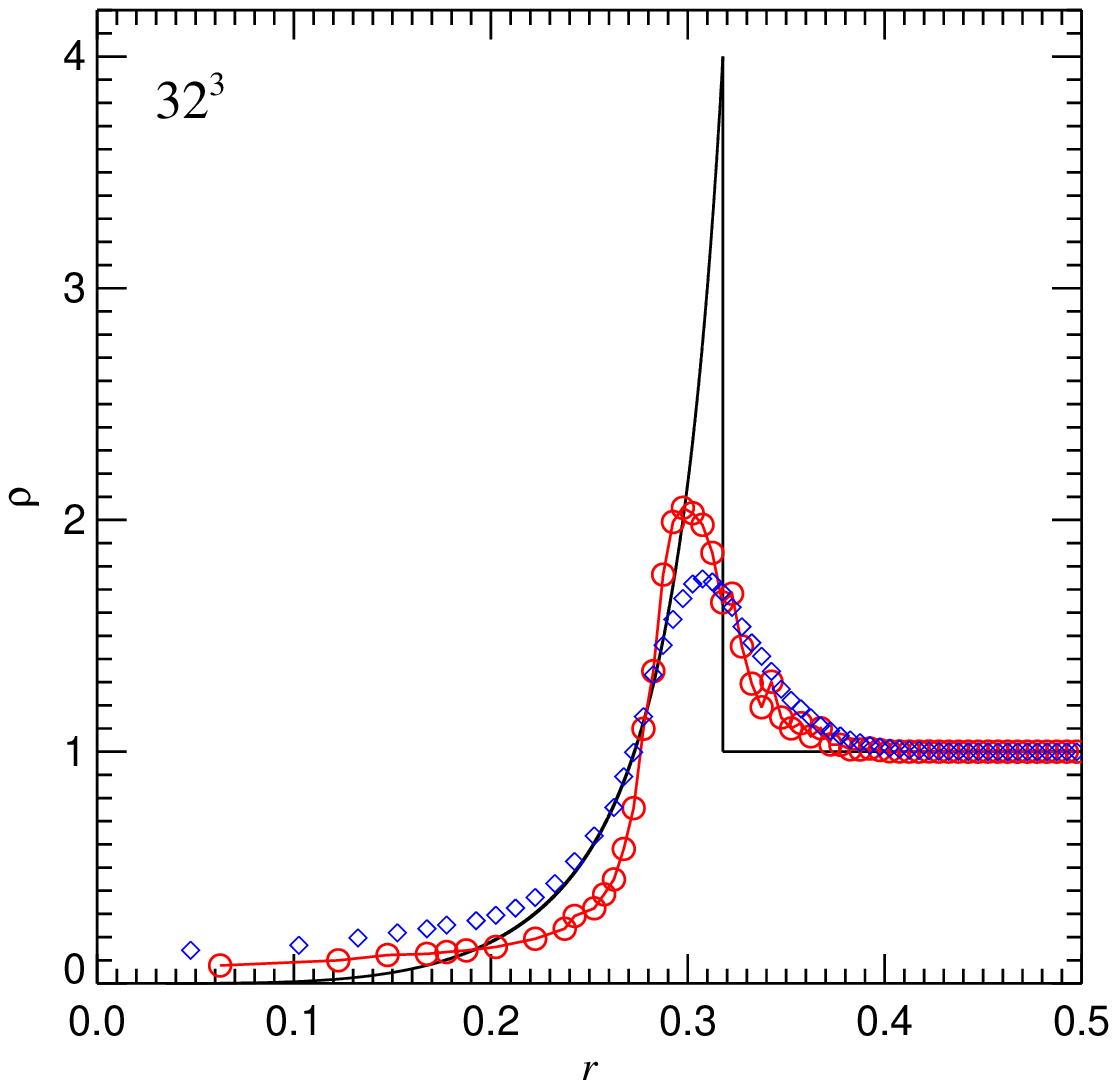}\\
\includegraphics[width=0.4\textwidth]{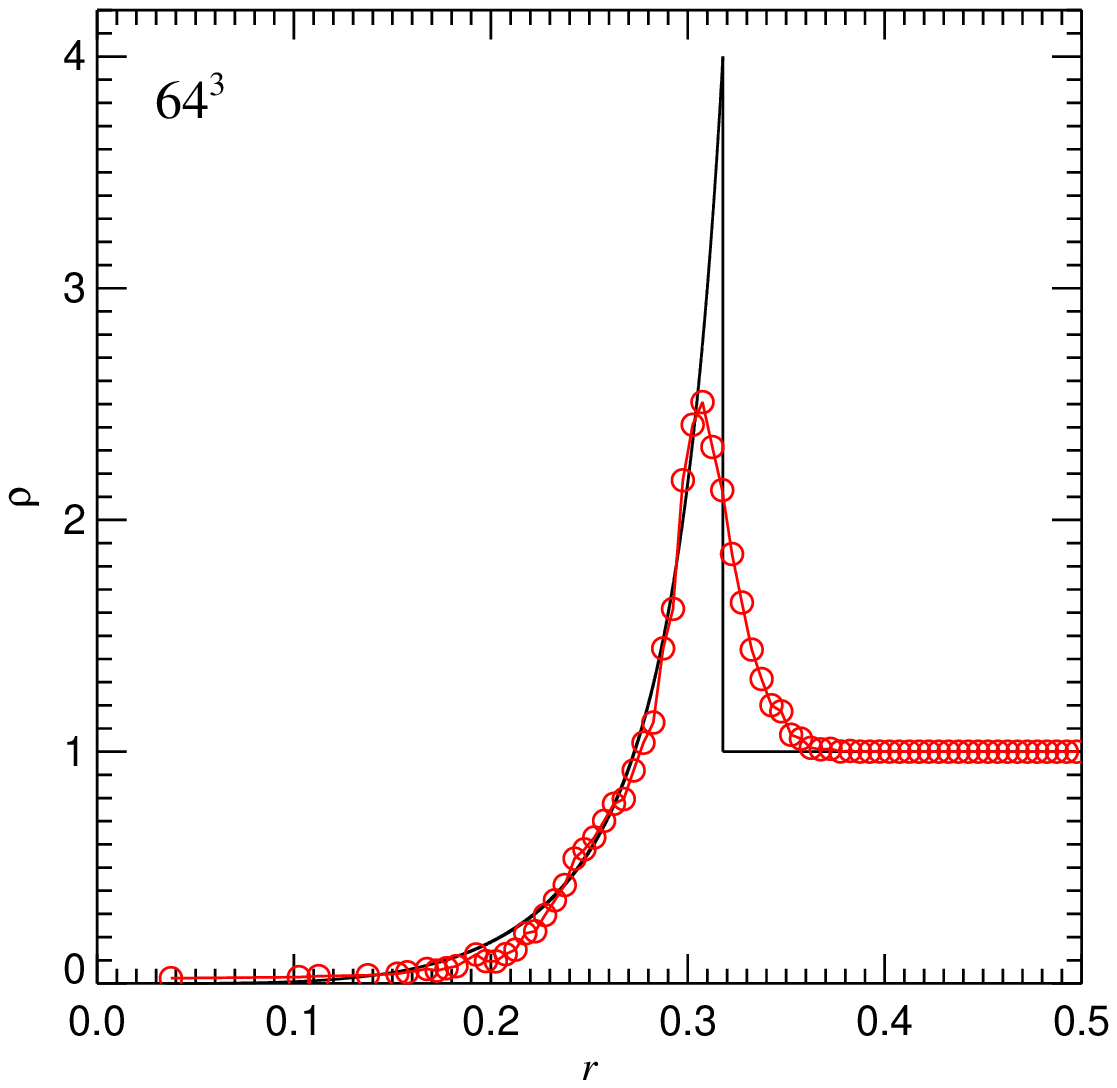}\\
\includegraphics[width=0.4\textwidth]{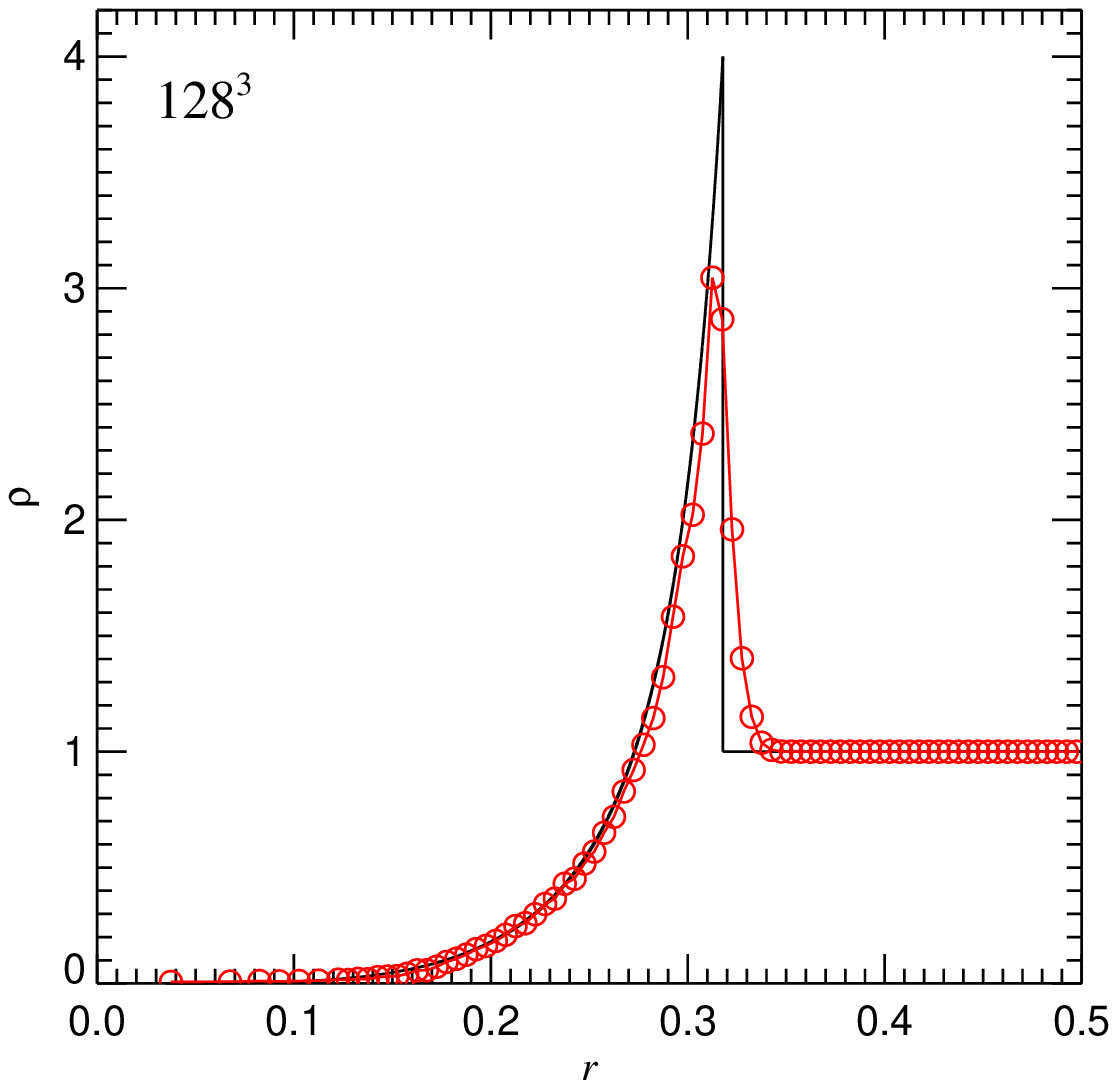}
\vspace{-4mm}
\caption{Sedov-Taylor point explosion problem, {\bf calculated in 3D with the
  basic VPH scheme. From top to bottom, we compare the radial density
  profile at $t=0.04$ with the analytical solution (solid line), 
calculated with $32^3$, $64^3$, or $128^3$ particles, as labeled. In the
top panel, we also compare the VPH results (circles) with SPH (diamonds).}
\label{Sedov_plot}
}
\end{center}
\end{figure}

\begin{figure}
\begin{center}
  \includegraphics[width=0.46\textwidth]{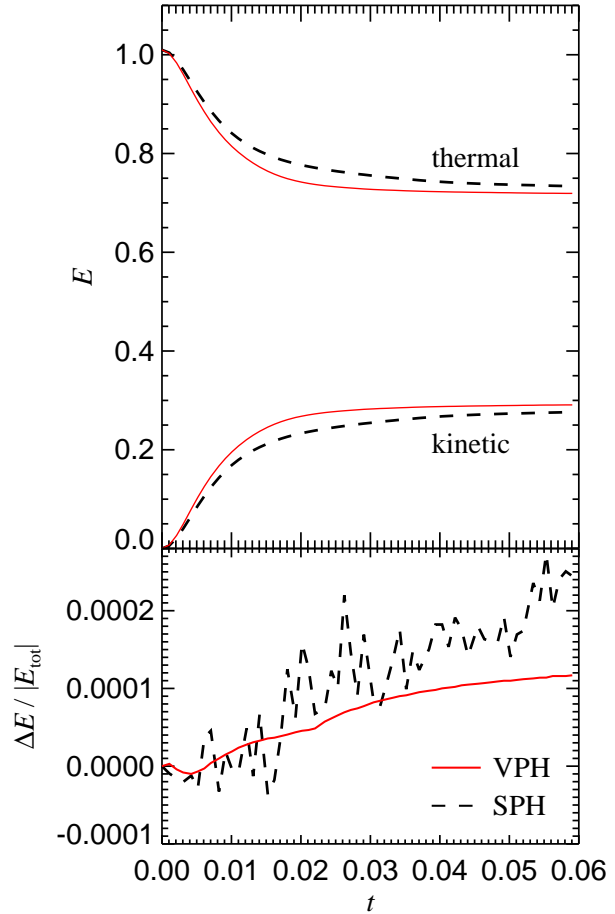}
\caption{Time evolution of the thermal and kinetic energies in the
  Sedov-Taylor point explosion problem, simulated with SPH (black
  dashed line) and the Voronoi scheme with shape correction forces
  (red solid line). The bottom panel compares the total energy error
  in the two schemes.
\label{Sedov_plot_energy}}
\end{center}
\end{figure}

In Figure~\ref{Sedov_plot}, we show the radial density profile at
$t=0.04$ {\bf for different resolutions corresponding to $2\times 32$,
  $2\times 64$ and $2\times 128$ particles}, and compare to the expected
analytic solution. The expected solution is captured reasonably well,
with a similar quality as in SPH codes, {\bf which is shown for
  comparison in the low-resolution case. The VPH result converges nicely
  to the analytic solution as the resolution is increased.} In
particular, the shock location is well reproduced, albeit with a small
pre-shock increase of the density.  When shape correction forces are
added as described in Section~\ref{SecShapeCorrect}, only negligible
differences in the overall quality of the result are found, but the
scatter around the azimuthally averaged solution is reduced. {\bf In
  Figure~\ref{Sedov_plot_energy}, we check the energy conservation in
  the blast wave problem, both for the VPH and SPH simulations at the
  $32^3$ resolution. The energy error is negligibly small, as
  desired. The high-frequency oscillations in the total energy of the SPH
  run stem from the small but finite jumps of the SPH smoothing lengths
  from timestep to timestep, an effect that is absent in the VPH
  simulation.}

\subsection{Kelvin-Helmholtz instabilities} \label{KHI}

Kelvin-Helmholtz (KH) instabilities occur in regions of strong shear,
which is especially common at contact discontinuities between two
fluid phases. An initially small transverse perturbation along the
interface becomes amplified and grows in linear theory according
to $\propto \exp(t/t_{\rm KH})$. After a few characteristic 
timescales
\begin{equation}
 t_{\rm KH} =  \frac{\rho_1 + \rho_2}{2\, k\, v     \sqrt{\rho_1  \rho_2}},
\end{equation}
an initially wave-like perturbation becomes large and non-linear,
developing the typical KH-rolls.  Here $\rho_1$ and $\rho_2$ are the
densities of the two media, $v$ is the velocity jump parallel to their
common interface, and $k = 2\pi/\lambda_{\rm pert}$ is the wavenumber
of the perturbation with wavelength $\lambda_{\rm pert}$. For an ideal
gas, all wavelengths are unstable, and the smallest wavelengths grow
fastest. 

The KH instability is especially important for the development of
turbulence, and is thought to play a prominent role in stripping and
mixing processes occurring during galaxy formation. Recently, a number of
studies have pointed out that standard SPH has problems to correctly
capture the KH instability when the initial conditions contain sharp
density gradients \citep{Agertz,Price2008}. In certain cases, the
instability is suppressed completely and does simply not grow. This can
in part be understood in terms of the surface tension effect present in
SPH, as described earlier, because surface tension suppresses the growth
of KH instabilities below a critical wavelength \citep{Landau1966}.
Furthermore, the asymmetric particle density at the interface causes a
rearrangement of the points in SPH, such that a `gap' in the sampling
appears that causes relatively large errors in the pressure forces at the
interface \citep{Agertz}.

\begin{figure*}
\begin{center}
\includegraphics[width=1.0\textwidth]{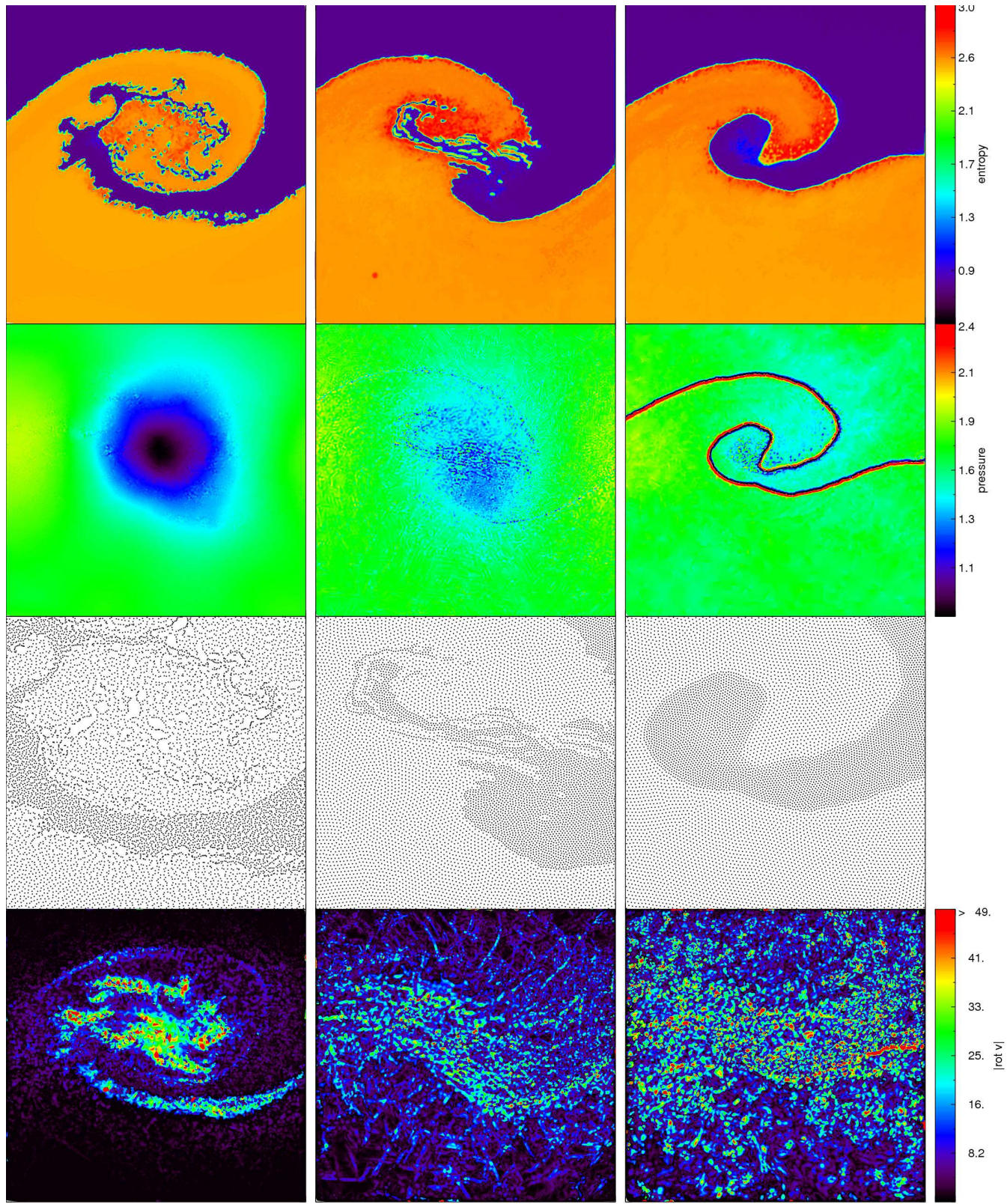}
\end{center}
\vspace{-26mm}
\caption{Simulations of the KH-instability with different
  particle-based methods. The left column shows the plain Voronoi
  scheme, the middle column Voronoi with PPO, and the right column
  SPH. From top to bottom, we show maps of specific entropy, maps of
  the pressure, the point distribution, and maps of the differential
  rotation $\nabla \times \vec{v}$. The maps show $x=[0.18,0.58],y=[0,0.4]$ of the
  periodic simulation domain, while for clarity the particle
  distribution is shown only for $x=[0.28,0.48],y=[0.1,0.3]$.
\label{KHI_chart}
}
\end{figure*}

It is therefore very interesting to test how well the VPH approach
does in this respect. Since VPH does not exhibit a surface tension
effect, it offers the prospect of a better treatment of the KH
instability. In Figure~\ref{KHI_chart}, we show results for a KH test
calculation, carried out with different particle-based hydrodynamic
schemes.  Our two-dimensional initial conditions consist of
$\gamma=5/3$ gas with density $\rho_1=2$ in the stripe $|y-0.5|<0.25$,
moving to the right with velocity $v_1=0.5$, and of gas with density
$\rho_2=1$ and velocity $v_2=-0.5$ in the region $|y-0.5| \ge
0.25$. The pressure was initialized everywhere to $P=2.5$, and a
periodic domain of unit length on a side was used. In total $261760$
points sample the gas distribution, with a mean spacing of $0.0023$
for the low-density gas, and $0.0017$ for the high density gas. Hence
the two phases were represented with approximately equal mass
particles. Note that the initial discontinuity was imposed as a
perfectly sharp jump in these initial conditions, following previous
studies of this problem. We remark however that it is somewhat
questionable whether such sharp jumps are not introducing an
inconsistency with the basic premises of SPH calculations, which can
only represent smoothed density fields.  In order to seed an initial
perturbation, we imposed a vertical
perturbation on the y-positions of the form
\begin{equation}
\delta y(x) = a_0\, \sin(4\pi x/L),
\end{equation}
where $L$ is the boxsize, and $a_0=0.006$ is the amplitude of the
initial perturbation.

\begin{figure}
\vspace{-2mm}
	\includegraphics[width=0.45 \textwidth]{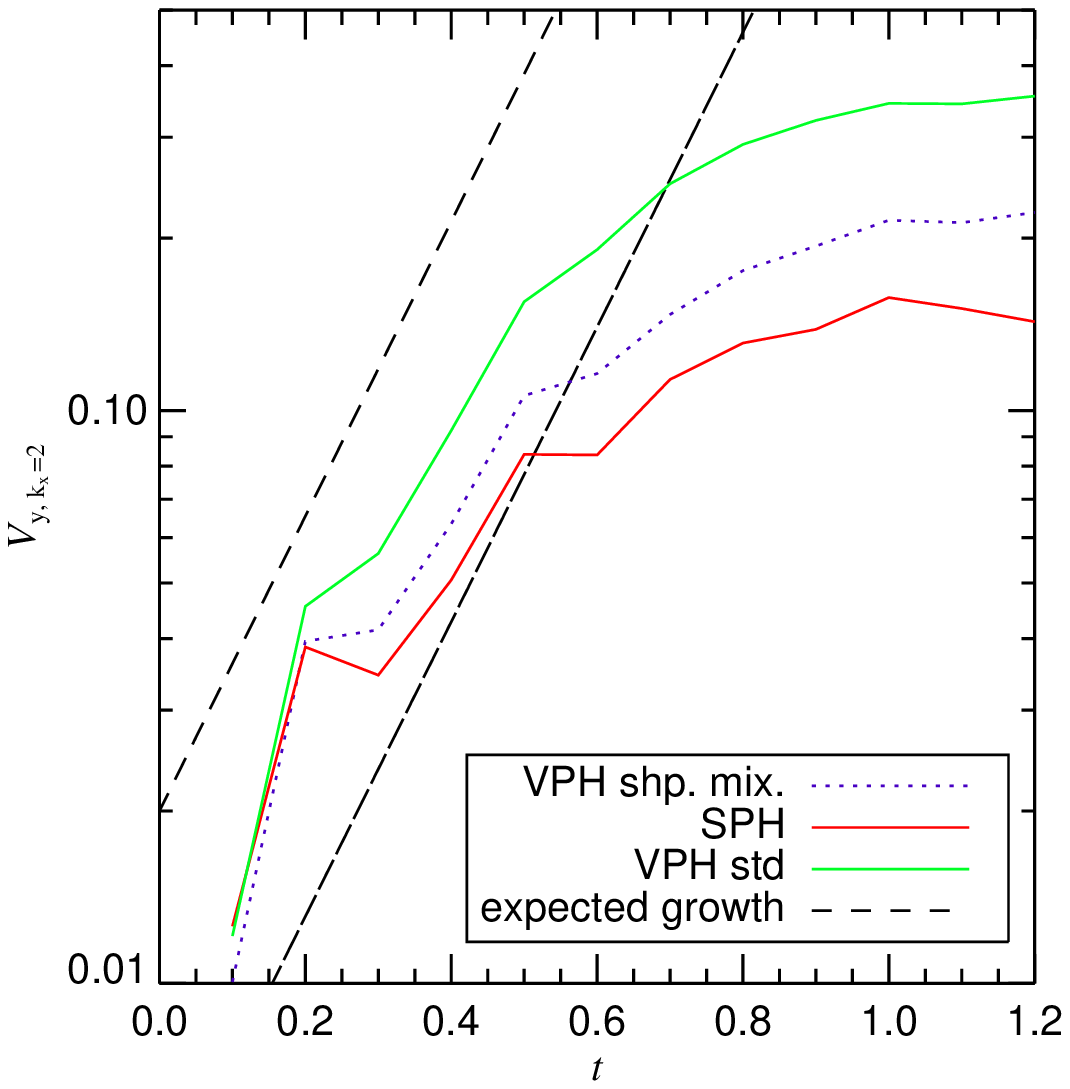}
\caption{Growth rate of the KH-instability for the same initial
  conditions as in Figure~\ref{KHI_chart}.  We show the amplitude of
  the seeded mode in the velocity field, measured by
  Fourier-transforming the $v_y$ field. We give results for standard
  VPH (green), for the Voronoi scheme with additional shape correction
  forces and the prescription for mixing discussed in
  Section~\ref{Mixing_Musings} (dotted blue), and for SPH (red). The
  dashed black lines indicate the slope expected for an exponential
  growth of the instability according to linear theory.  }
\label{KH_growth_rate}
\end{figure}

The three columns of Figure~\ref{KHI_chart} compare the results for
the ordinary VPH scheme (left), the VPH method with the additional
ordering viscosity of the PPO scheme (middle), and SPH (right), at
time $t=1.2$. From top to bottom, we show specific entropy maps,
pressure maps, the particle distribution, and vorticity maps.  All
maps were here generated by linearly interpolating a Delaunay
tessellation of the points, allowing to extend the points' properties
as read from the simulation files to continuous fields. In addition to
the maps, we show in Figure~\ref{KH_growth_rate} the growth of the
instability until time $t=1.2$, quantified in terms of the amplitude
of the seeded velocity mode as measured in the Fourier-transformed
$v_y$-field.

We see right away that the VPH scheme captures the KH instability
best. Its primary KH billow has evolved furthest, and it triggered the
growth of smaller-scale secondary billows. In contrast, the SPH result
shows only an anemic growth of the instability.  In the SPH pressure
map a strong pressure anomaly is visible at the interface.  This
surface effect effectively suppresses all small-scale
KH-instabilities, and the two phases stay separate because the
associated surface tension suppresses a breaking up of the interface.
As a result, the instability cannot cascade down to smaller scales.
The PPO scheme shown in the middle column lies literally in the middle in
this respect. The growth of the primary KH mode is very similar to
that found in the VPH scheme. However, the additional viscosity
introduced in this scheme to produce highly regular cells
substantially attenuates the growth of secondary small-scale KH
instabilities. The same effect can be seen Figure~\ref{KH_growth_rate}
for the case with shape correction terms, whereas additional heat diffusion 
(according to Section~\ref{Mixing_Musings}) 
does not affect the growth rate of the excited mode
in this simulation. We note however that the smaller growth depends on
the strength of the additional ordering forces that are invoked. The
result shown here was calculated with $\kappa =1$, which we consider the
maximum that one may ever want to use. For more reasonable smaller
values, intermediate results that are close to that of the plain VPH
scheme are obtained.

Interestingly, the vorticity $\nabla \times \vec{v}$ in the ordinary
VPH scheme is clearly largest overall, especially for the larger modes
and in the central
region of the primary KH billow. Here the rotation is so fast that the
pressure shows a noticeable depression, which counteracts the 
centrifugal forces from the rotation with pressure
gradients. Unfortunately, the ability of the pure Voronoi simulation to
sustain vortices comes at the expense of a larger particle
irregularity. As the enlargement with the particle distribution shows,
the particles tend to form
Voronoi cells with quite high aspect ratios, reducing the
accuracy of the gradient estimates and requiring relatively large settings
for the artificial viscosity to prevent interparticle penetrations.
In contrast, the PPO variant of the Voronoi scheme produces highly
regular particle spacings, and in this respect resembles SPH.
However, in this example calculation with $\kappa =1$, the scheme then
develops effectively a much higher intrinsic shear viscosity, which
tends to transform the differential rotation into a rotation of numerous
ordered domains. The shear of these domains shows up in the
differential vorticity maps in a distribute fashion. Once the
damping of the differential rotation gets too strong, the primary
vortex is affected as well, as seen in the reduction of the central
pressure gradient in the PPO scheme.

\begin{figure}
\vspace{-2mm}
	\includegraphics[width=0.4\textwidth]{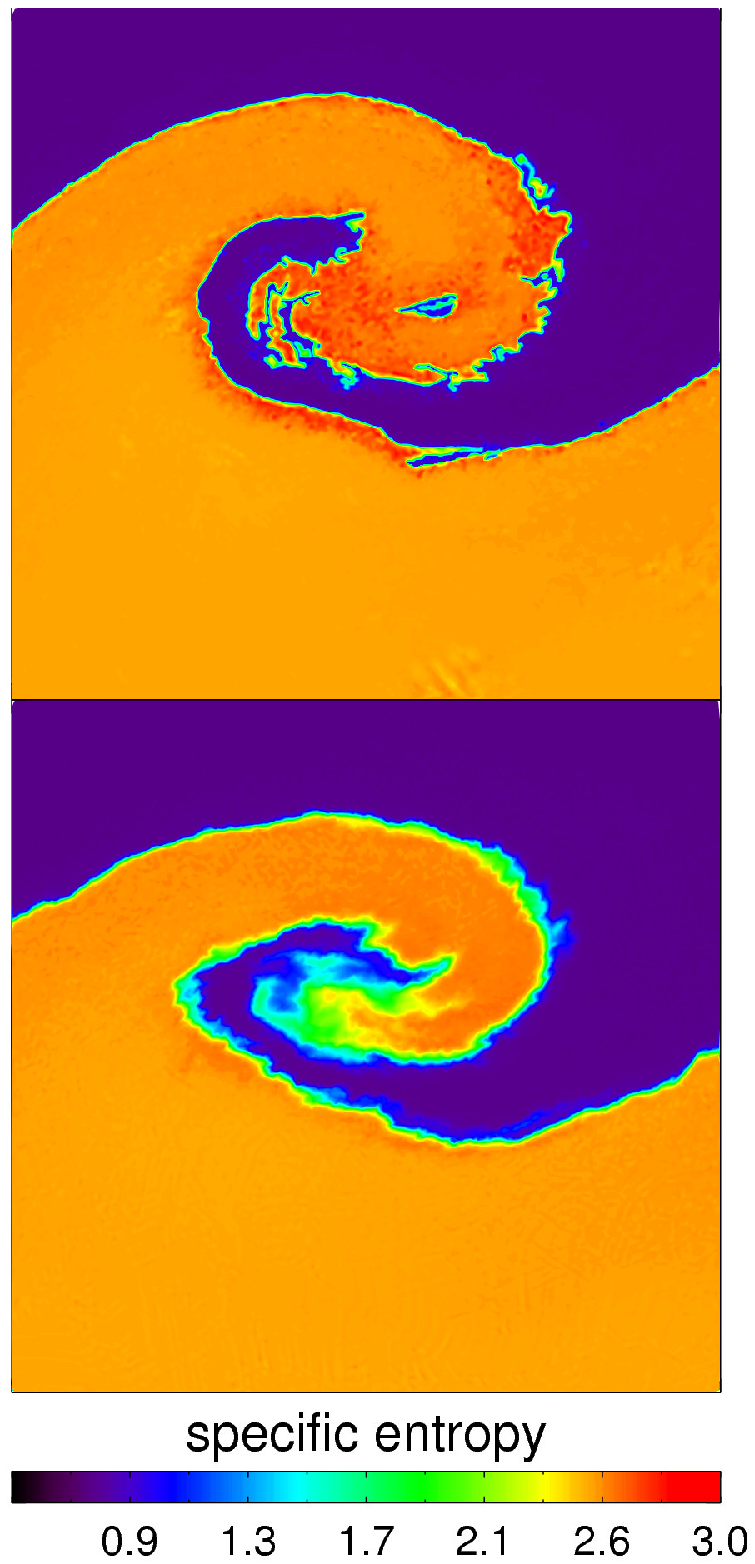}
\caption{KH-instability test simulated with the Voronoi scheme with
  additional shape correction forces, based on the same initial
  conditions as in Figure~\ref{KHI_chart} and again at time $t=1.2$.
  The maps show the entropy distribution without (top) and with
  (bottom) additional heat diffusion terms to model subresolution
  mixing, as described in Section~\ref{Mixing_Musings}.  
}
\label{KH_with_diffusion}
\end{figure}

Finally, we test the shape correction forces discussed in
Section~\ref{SecShapeCorrect}, and our scheme for the treatment of
subresolution mixing introduced in Section~\ref{Mixing_Musings} with
the same initial conditions. We show
in Fig.~\ref{KH_with_diffusion} the resulting entropy maps of two
further calculations of the KH instability test.
In both panels, we show
runs of the VPH scheme that make use of the shape correction forces
derived from the Lagrangian, with $\beta_{0} = 1.2$ and $\beta_{1} =
0.1$. In the bottom panel, we have in addition activated the artificial
heat conduction due to local shear with $\chi =0.25$, which models mixing of the fluids at
the scale of the resolution.  The two variants are qualitatively
similar, but the artificial heat conduction has clearly washed out the
sharp discontinuity in the entropy at the fluid interface. This
resembles more closely the results of mesh-based finite-volume hydro
codes. Compared to the results of pure VPH in Figure~\ref{KHI_chart}, we see that
the scheme with shape correction forces has also an effectively enlarged
viscosity, quite similar to the PPO approach.  Overall, it is 
clear that even without a subresolution mixing model the Voronoi based
particle hydrodynamics does significantly better in the KH test than
standard SPH.

\subsection{The `blob test': mass loss of a gas cloud in a supersonic wind}

A challenging test problem for hydrodynamical codes has been proposed
by \citet{Agertz}. The setup consists of an overdense spherical cloud
in pressure equilibrium with the surrounding hot medium. This
background gas is given a large velocity, so that  the
cloud feels it as a supersonic head wind. The test is motivated by
astrophysical situations such as the stripping of gas out of the halos
of galaxies as they fall into larger systems. It is a
three-dimensional problem that involves many different non-linear
hydrodynamical phenomena, including shocks, Kelvin-Helmholtz
instabilities, mixing, and the generation of turbulence.  Because of
this complexity, an analytical solution for the problem is not known.
The general expectation is that the wind will compress the cloud,
accelerate it, and strip some of its gas by developing fluid
instabilities as it streams past the cloud.

Interestingly, in the test calculations of \citet{Agertz}, substantial
differences were found in the mass loss rates of the cloud when
calculated with Eulerian mesh codes and with SPH. Whereas the mesh
codes led to an eventual complete destruction of the cloud, the mass
loss rate was in general smaller in SPH, such that some cloud material
still remained once the partially destroyed cloud was accelerated to
the wind speed and the mass loss stopped. Given these qualitatively
different outcomes, it is interesting to test how the VPH scheme
performs on this problem.

We used the same initial conditions as employed by \citet{Agertz}, in
the version with $10^6$ particles.  The setup consists of a periodic
box with extension $[0,2000]\times
[0,2000]\times[0,8000]\,\mathrm{kpc}$. The background `wind' gas has
density $\rho_{\mathrm{wind}}=4.74\times 10^{-34}\,{\rm g\, cm^{-3}}$,
temperature $T_{\rm wind}=10^7\,{\rm K}$, and a velocity
$v_{\mathrm{wind}}=(0,0,1000)\,\mathrm{km\,s^{-1}}$. The cloud has a
radius $R_{\mathrm{cloud}}=197\:\mathrm{kpc}$, and is placed
initially at
$\vec{r}_{\mathrm{cloud}}=(1000,1000,1000)\,\mathrm{kpc}$. Its density
is 10 times higher than the background, $\rho_{\mathrm{cloud}}=10
\,\rho_{\mathrm{wind}}$, while at the same time being 10 times colder,
$T_{\rm cloud}= 10^6\,{\rm K}$. This yields a sound speed of
$c_{\mathrm{wind}}=371\, \mathrm{km\,s^{-1}}$ for the wind, and an
expected characteristic timescale of order $\tau_{\rm KH}\simeq
2\,{\rm Gyr}$ for the development of large KH instabilities in the
shear-flow around the cloud.

\begin{figure}
\vspace{-2mm}
\includegraphics[width=0.42\textwidth]{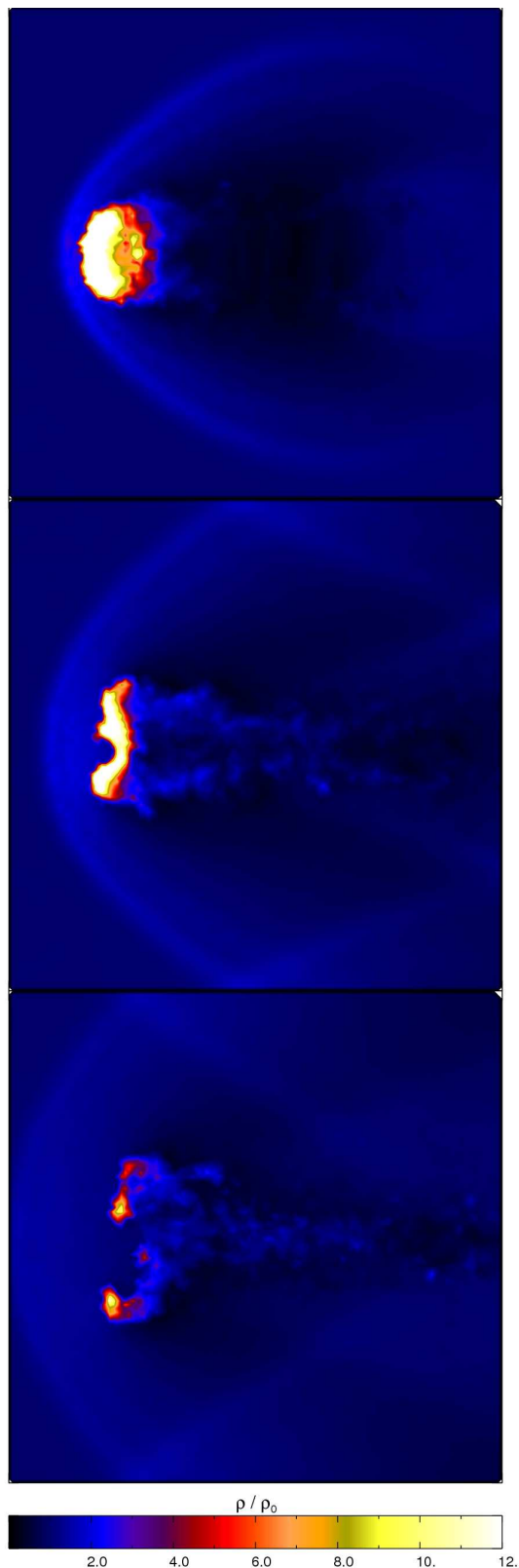}
\vspace{-10mm}
\caption{Time evolution of the density for a gas cloud in a
  supersonic wind. From top to bottom, we show density maps normalized
  to the initial wind density at times $t
  =0.75\, \tau_{\mathrm{KH}}$, $t=1.5\, \tau_{\mathrm{KH}}$, and
  $t=2.25 \tau_{\mathrm{KH}}$ in the central plane of the simulation
  box. Here the standard Voronoi scheme with $10^6$ particles was
  used. }
\label{BLOP densities}
\end{figure}

In Figure~\ref{BLOP densities}, we show the time evolution of density
slices through the central plane of the simulation box, calculated
with our new Voronoi scheme. In agreement with both grid-based and SPH
codes, a bow shock is produced ahead of the cloud.  The cloud gets
compressed and accelerated under the ram pressure of the wind, and the
wind that streams past the deformed and slowly accelerating cloud
induces Kelvin-Helmholtz instabilities at its surface which strip 
material and produce a turbulent wake. At the stagnation point of the
flow in front of the cloud, the pressure eventually breaks through
along the axis of symmetry.  The resulting ``smoke ring'' is then
still exposed to Kelvin-Helmholtz induced turbulence while it is being
accelerated to the velocity of the background stream. Due to the
periodic boundary conditions we note that the bow shock extends past
the domain boundary and then back inwards again from the other side,
reaching the cloud at about $t\approx 3\, \tau_{\rm KH}$. This leads to a
recompression of the remainder of the cloud which can temporarily raise
the number of particles that are still counted as cloud members.

\begin{figure}
\vspace{-6mm}
\hspace{-6mm}
	\includegraphics[width=0.5\textwidth]{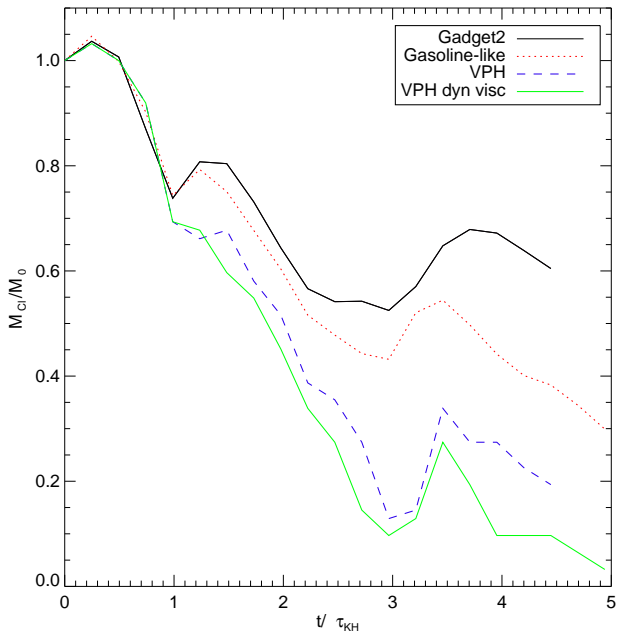}

\vspace{-4mm}
\caption{Mass loss of a gas cloud in a supersonic wind (the `blob
  test'), simulated with $10^6$ particles and different hydrodynamical
  schemes. We show the remaining mass associated with the cloud
  (particles that fulfill $\rho > 0.64 \: \rho_{\mathrm{cloud}}$ and
  $T < 0.9 \, T_{\mathrm{wind}} $) as a function of time (in units of
  $\tau_{\rm KH}$) for different hydrodynamical schemes. The different 
  colors refer to the SPH codes {\small GADGET2} (black), 
  modified {\small GADGET2} with an {\small GASOLINE}-like integration
  scheme (red), standard VPH (green), VPH with dynamic viscosity (see
  {\bf{Equation~\ref{dyn_visc_eq}}} (light blue) and VPH with heat 
  diffusion according to Section~\ref{Mixing_Musings} (dark blue)
}
\label{Agertz_massloss}
\end{figure}

The mass loss as a function of time is displayed in
Figure~\ref{Agertz_massloss}, for different particle-based
hydrodynamical schemes. We follow \citet{Agertz} and consider a
particle to be still part of the cloud when its density is still
larger than $\rho > 0.64 \: \rho_{\mathrm{cloud}}$, and its
temperature fulfills $T < 0.9 \, T_{\mathrm{wind}} $. We show results
for four different calculations in total. The standard VPH scheme is
shown in blue. Interestingly, it leads to a complete destruction of
the cloud at time $t\approx 3\, \tau_{\rm KH}$, a result which is
actually surprisingly close to the high-resolution mesh-based
calculations reported in \citet{Agertz}. On the other hand, the two
SPH-based results (shown in red and black) calculated with the {\small
  GADGET2} code do not result in a destruction of the cloud. Instead
at time $t= 5\, \tau_{\rm KH}$, still about half the mass of the
original cloud can be characterized as residual cloud material. This
is even slightly larger than what was reported by \citet{Agertz} for
the {\small GASOLINE} code.  We have however found that the
formulation and the strength of the artificial viscosity can influence
this result significantly. Also, we confirmed that integration of the
entropy as independent thermodynamic variable (which is the default in
{\small GADGET2}) results in less stripped material than when the
thermal energy is integrated as done in {\small GASOLINE}. This is
however probably largely a result of the particular initial conditions
used here; the contact discontinuity in the ICs of \citet{Agertz},
that we employ here, has been relaxed using the traditional SPH
formulation of {\small GASOLINE}, creating a pressure blip. When this
is then used to initialize the entropies integrated in {\small
  GADGET2}, a spurious entropy blip is created that further amplifies
the initial sampling `gap'.

When strong shape correction forces are introduced into the VPH
formalism, we find an intermediate result between plain VPH and
SPH. In this case, a small 10\% remnant of the cloud remains at time
$t= 5\, \tau_{\rm KH}$. This is consistent with our earlier findings
for the KH instabilities. While standard SPH can be expected to
suppress the KH-instabilities significantly at the cloud interface, it
tends to underestimate the rate of stripping. Our new VPH method does
not show this problem, but if viscous forces are introduced that
guarantee very regular mean particle separations, some small-scale
suppression of fluid instabilities can be reintroduced.

We finally note that the use of a time variable viscosity as presently
implemented in our code has not changed the mass-loss curves
significantly. The reason is that the relevant particle viscosities are
pushed to a large value as they pass through the bow shock, and stay at
large values in the complicated flow around the cloud surface.  Only at
late times the mass loss tends to become faster as a result of the
effectively lower viscosity.

\subsection{Gravitational collapse of a gas sphere}

Finally, we consider a three-dimensional problem with self-gravity,
the so-called `Evrard-collapse' \citep{Evrard}. It consists of an
initially cold gas cloud, with a spherically symmetric density profile
of $\rho(r) \propto {1}/{r}$. The total mass, outer radius and
gravitational constant are all set to unity, $M=R=G=1$, and the
initial thermal energy per unit mass is set to $u=0.05$. In this
configuration the sphere is significantly underpressurized. It hence
collapses essentially in free-fall, until it bounces back at the
centre, with a strong shock running through the infalling
material. The sphere then settles into a new virial equilibrium.  As
this problem involves large conversions of potential gravitational
energy into kinetic energy and thermal energy (and back), as well as
strong shocks, it is a challenging and useful test for hydrodynamic
codes that are applied to structure formation problems. For this
reason, it has been widely used as a test for a number of SPH codes
\citep[e.g.][]{Evrard,Hernquist1989,Steinmetz1993,Dave1997,gadget1,Wadsley2004,
  gadget2}.

\begin{figure}
\vspace{-6mm}
	\includegraphics[width=0.45\textwidth]{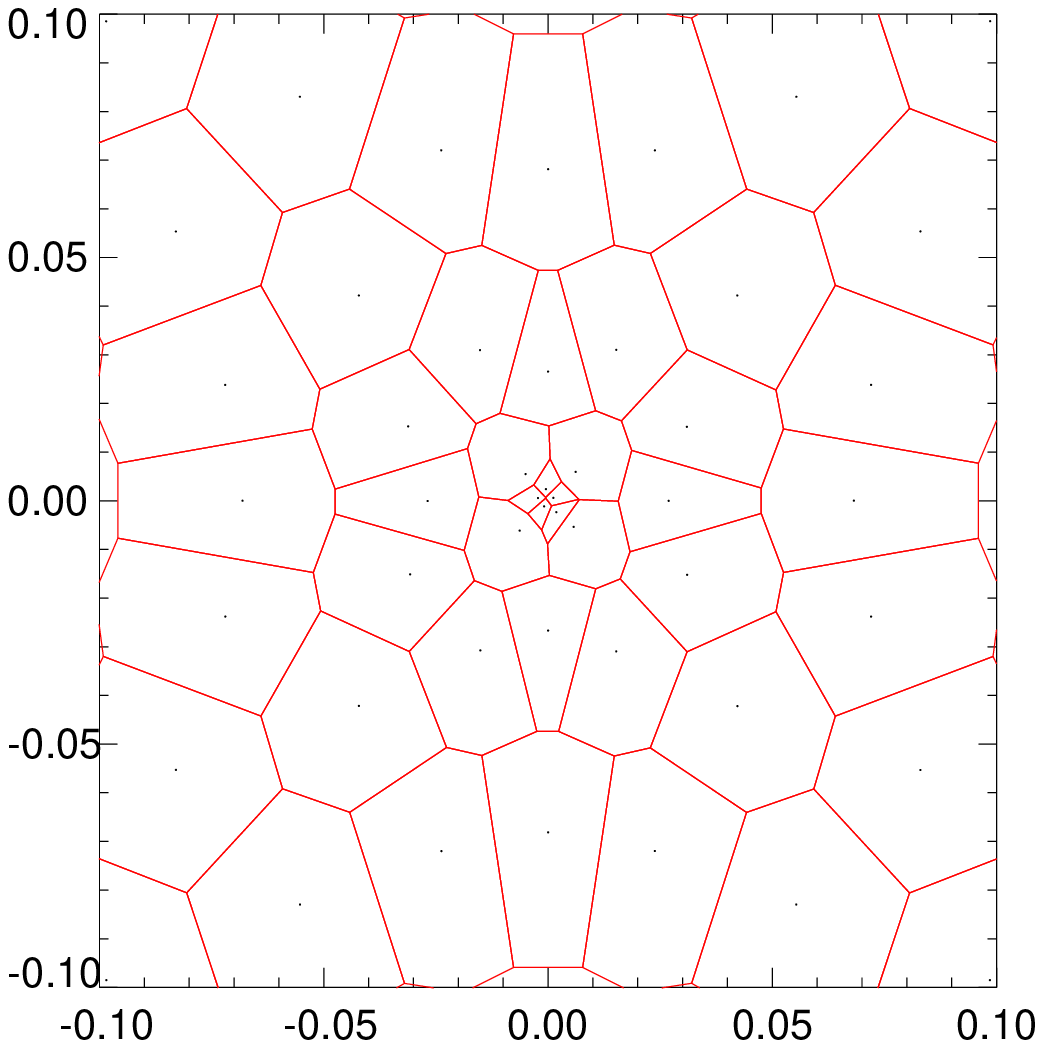}
	\includegraphics[width=0.45\textwidth]{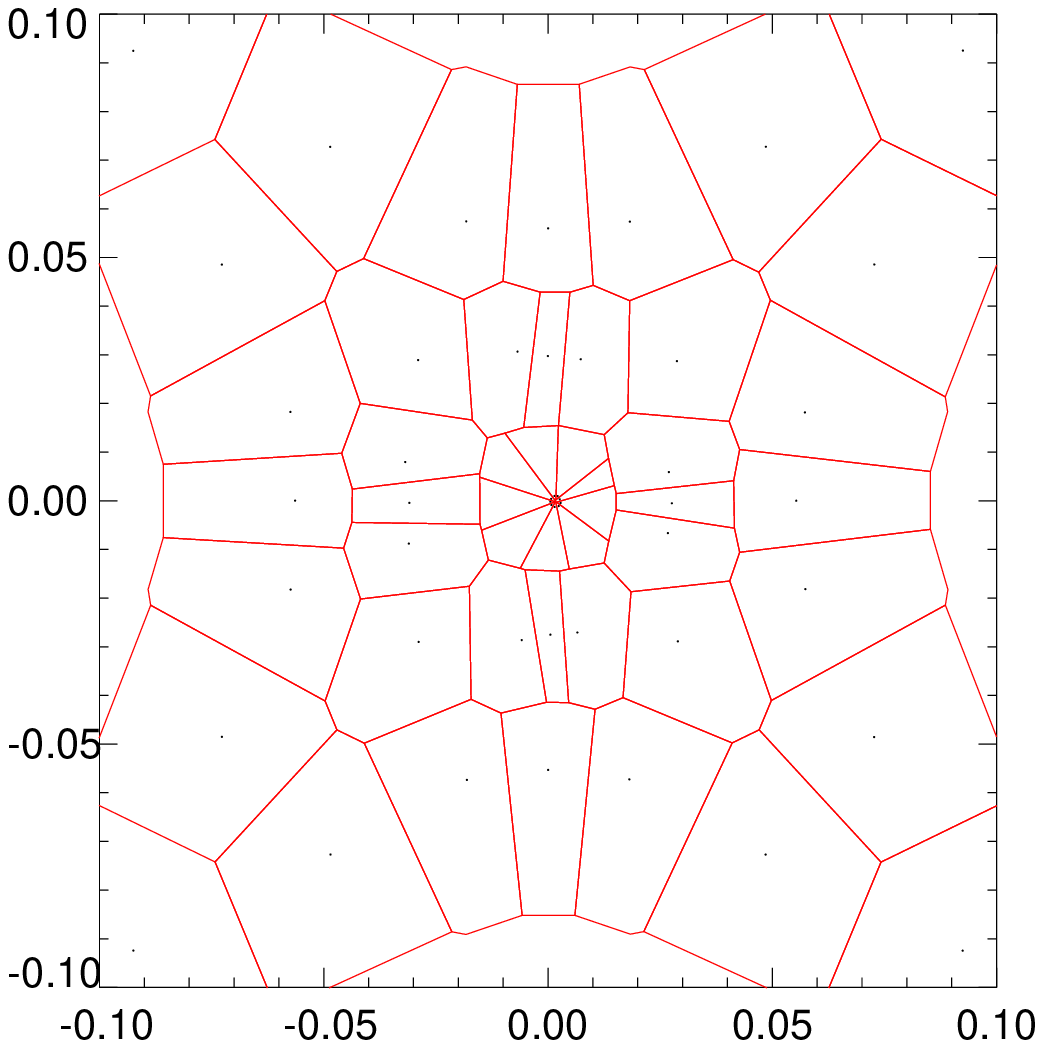}
\vspace{-4mm}
\caption{Cell regularity without (bottom) and with cell regularization
  forces (top) in a simulation of a sphere of gas collapsing under
  self-gravity (Evrard collapse problem). In the bottom panel, the group
  of points clusters together close to the origin under their mutual
  gravitational attraction, producing a quite irregular mesh there.
  This effect is prevented when additional shape correction forces are
  invoked, as shown in the top panel.
\label{FigMeshEvrard}
}
\end{figure}

For our test we create a realization of the sphere by stretching a
Cartesian grid appropriately, such that the desired initial density
profile is obtained.  Because the Voronoi scheme needs to tessellate a
well-defined total volume, we cannot impose vacuum boundaries in the
same way as in SPH. Instead, we embed the sphere in a box and use a
background grid of particles {\bf with extremely rapidly falling density
  profile outside of the sphere}, such that the total mass in the
background can still be ignored in the evolution of the system. We
calculate the gravity with the same tree algorithm used in the {\small
  GADGET2} code, simply using the N-body approach with the masses and
positions of the VPH particles. It turns out however that particles
sometimes tend to pair up under their pairwise gravitational forces in
the plain VPH scheme. This is related to the same defect discussed in
the context of Figure~\ref{FigDegeneracy}. If two particles are very
close to a Voronoi wall, they can be moved still closer together without
increasing the hydrodynamic pressure force, so that a residual
gravitational attraction (if not damped out by the gravitational
softening) can move the particles very close together, with problematic
consequences for the stability and accuracy of the scheme.  A
time-dependent gravitational softening \citep{Price2007}, where the
softening is somehow tied to the size of the cell associated with a
particle, may ease the problem, but is unlikely to cure it completely.
However, the shape correction forces we introduced in
Section~\ref{SecShapeCorrect} can nicely solve this problem. This effect
is illustrated in Figure~\ref{FigMeshEvrard}, where we show the central
mesh geometry for a two-dimensional version of the Evrard collapse
problem.  In the following we consider therefore a calculation of the 3D
Evrard collapse with our usual choice of $\beta_{0} = 1.2$ and
$\beta_{1} = 0.1$.

\begin{figure}
\begin{center}
\resizebox{8cm}{21cm}{\includegraphics{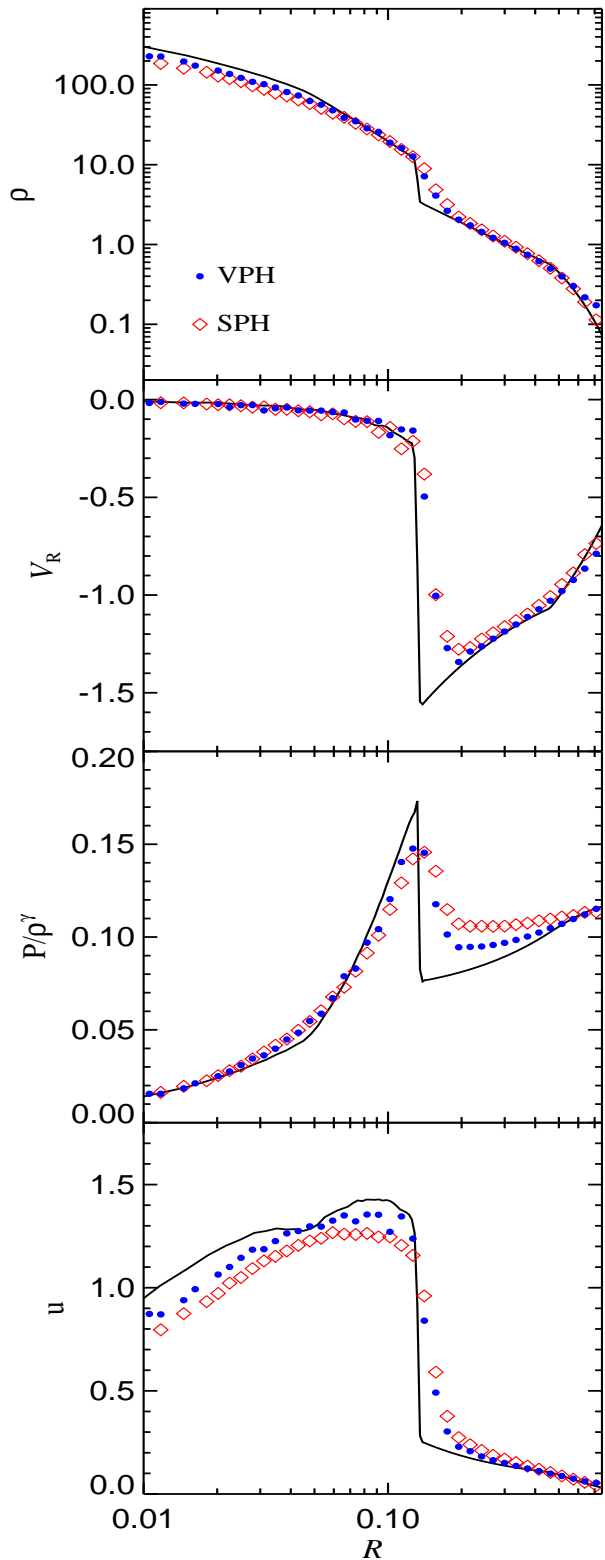}}
\caption{Evrard collapse at time $t=0.8$ simulated with SPH (red diamonds)
  and the Voronoi scheme with shape correction forces (blue circles). The black
  line shows the results of a 1D high-resolution PPM calculation of
  the problem provided to us by \citet{Steinmetz1993}. From top
  to bottom, we show radially averaged profiles of gas density, radial
  gas velocity, specific entropy, and internal energy.}
\label{evrard}
\end{center}
\end{figure}

\begin{figure}
\begin{center}
\vspace{-5mm}
  \includegraphics[width=0.45\textwidth]{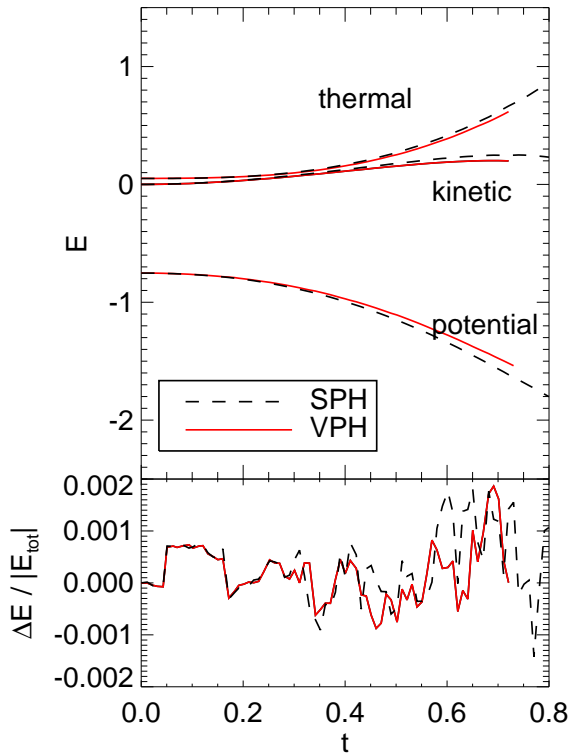}
\caption{Energy evolution for the ``Evrard collapse'' simulated with
  SPH (black dashed line) and the Voronoi scheme with shape correction
  forces (red solid line). The small fluctuations in the total energy
  arise primarily as a result of the finite accuracy of the tree code
  used to calculate self-gravity, and are of similar magnitude for
  both cases.}
\label{evrardBudget}
\end{center}
\end{figure}

In Figure \ref{evrard}, we show radial profiles of gas density, radial
velocity and entropy for the Evrard collapse at time $t=0.8$, calculated
with 24464 particles inside the initial sphere. We compare the VPH
result (shown by blue circles) to results obtained with SPH (red
diamonds), and compare these to results of a 1D high-resolution PPM
calculation of the problem (solid black line) provided to us by
\citet{Steinmetz1993}. {\bf We find that the collapse is essentially
equally well described with VPH as with SPH, with a slight advantage for
VPH, which more accurately resolves the central density and captures the
shock more sharply.  However, these differences lie well in the range of
changes one obtains for different viscosity prescriptions, and therefore
do not seem to be particularly significant.}  We also note that the extra
viscous forces needed to maintain a regular particle distribution in VPH
do not introduce any unphysical features in the solution. In particular,
the radial profile of the specific entropy shows no signs of extra
cooling or heating.

\section{Conclusions}  \label{SecConclusions}

We have discussed a new fluid particle model where the density is
estimated with the Voronoi tessellation generated by the particle
positions. Unlike in SPH, there is an auxiliary mesh, which adds
complexity to the scheme. However, the use of this fully adaptive mesh
offers a number of advantages. It offers higher resolution for a given
number of particles, since fluid features are not inherently smoothed
as in SPH. In fact, the tessellation techniques are probably close to
an optimum exploitation of the density information contained in the
particle distribution \citep{Pelupessy}. When even higher resolution
is needed the method could be extended by an adaptive particle
refinement or splitting technique, for example similar as suggested in
\citep{AREPO}.

As a result, contact discontinuities can be resolved with one cell, and
surface tension effects present in standard SPH across contacts with
large density jumps are eliminated. This has further implications for
the growth rate of fluid instabilities in inviscid gases. Furthermore,
the free parameters in the density estimate of SPH, involving both the
number of neighbours as well as the kernel shape, are eliminated, which
can be viewed as a good thing since the optimum values for them are not
known, and incorrect choices can invoke the well-known clumping
instability in SPH.

One somewhat problematic aspect of Voronoi-based particle hydrodynamics
is that the noise in the scheme is quite sensitive to the level of mesh
regularity. Flows with a lot of shear can readily develop
Voronoi meshes with points that lie close to the surfaces of their
Voronoi cells. In this case the noise in the gradient estimates
increases, and, more importantly, it becomes difficult to safely prevent
particle interpenetration, since closeness to a wall of the
tessellation always implies that there is a second point on the other
side of the wall which is also close, i.e.~in other words, that a close
particle pair is present.

Higher order density estimates might solve this problem, but they
would have to be introduced already into the Lagrangian, leading to
much more complicated equations of motion that may be intractable. We
therefore explored two different approaches for keeping the mesh
relatively regular. One is simply based on trying to formulate
additional artificial viscosity terms such that the viscosity tries to
make cells `rounder'. Whereas this shows some success, it does not
succeed in all situations, particularly in strong shear flows where
the artificial viscosity needs to be very low. A more radical approach
also explored is to add correction terms to the underlying fluid
Lagrangian with the aim to penalize strong deviations from regular
mesh geometries. Our goal was to impose small, non-dissipative
correction forces that maintain a proper mesh geometry. Thanks to the
Lagrangian formulation, the required form of the correction forces to
retain fully conservative behaviour can readily be derived, and the fluid
motion under these forces shows the desired properties. However,
if the correction terms are too large, one risks deviations from the
proper hydrodynamic solution.  Further experimentation will be
required to identify the optimum setting of these parameters.

The Voronoi-based fluid particle approach can be relatively seamlessly
integrated into an existing SPH code, provided a tessellation engine can
be added in an appropriate fashion. Other aspects of the physics (in
particular self-gravity, an additional collisionless component,
radiative cooling, star formation, and feedback processes) can be
treated in essentially identical ways as in SPH. This makes it possible
to readily apply Voronoi particle hydrodynamics to problems of interest
in cosmological structure formation.  In general, our first results
suggest that VPH is superior to SPH, albeit at much increased
complexity. However, it is at present still unclear whether it is
competitive with finite volume hydrodynamics carried out on a similarly
constructed Voronoi mesh, as realized in the {\small AREPO} code
\citep{AREPO}. To elucidate this point further, we are in the process of
carrying out galaxy collision simulations as well as cosmological
structure formation simulations with our new technique and will report
the results in forthcoming work.

\bibliography{DRAFT}
\bibliographystyle{mn2e.bst}


\onecolumn
\appendix

\section{Differential operators on Voronoi meshes}

In this Appendix, we collect some useful formulae for discretized
versions of differential operators on Voronoi meshes, such as the
gradient or the divergence, and we test the accuracy of the gradient
estimate as a function of the regularity of the Voronoi mesh.

\subsection{Gradient}
\label{app_grad}
\begin{figure}
\begin{center}
  \includegraphics[width=0.4\textwidth]{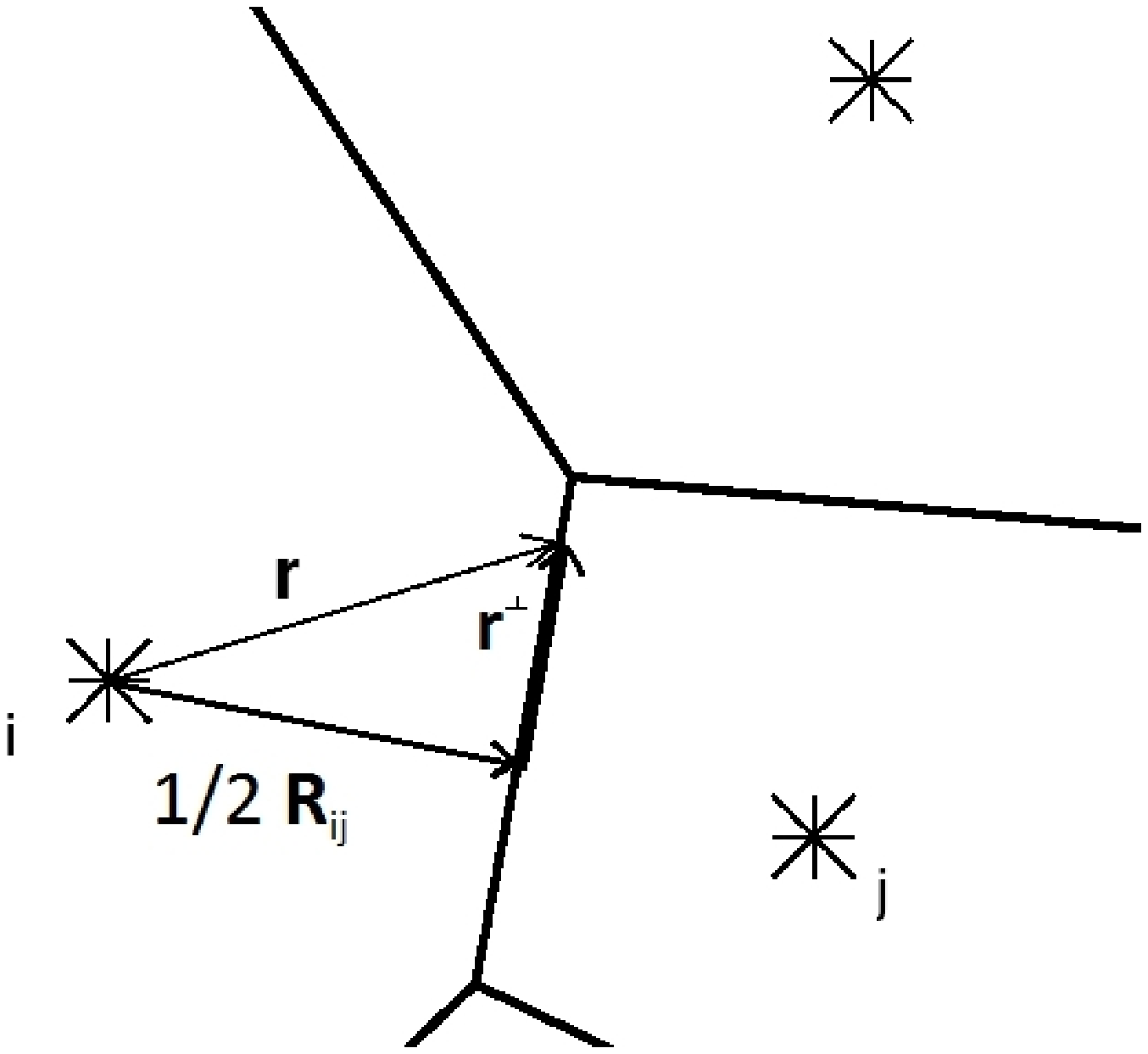}
\caption{Sketch of the integration variables 
\label{Sketch}
}
\end{center}
\end{figure}

The cell averaged gradient of any quantity $\phi$ can be estimated via
Gauss' theorem. One can use Gauss' theorem on a product of a constant 
vector times a scalar field and arrives at:
\begin{equation}
 \frac{1}{V} \int_{V} \vec{\nabla} \phi  \:  {\rm d}V= \frac{1}{V}
\int_{\partial V} \phi
 \: {\rm d}\vec{S},
\end{equation}
which can be used as one way to derive an estimate of the local gradient
by approximating the value of $\phi$ on the surface of a cell with the
arithmetic mean between the cell and its neighbours.  However, one can
also circumvent the problem of finding a proper value for $\phi$ on the
surface by using a different starting point.  
We now use Gauss' theorem on $ (\vec{1} \cdot \vec{r}) \, \vec{\nabla} \phi$, where
$ \vec{1}$ is the unit vector. This yields:
\begin{equation}
\int_V  \vec{\nabla} \phi {\rm d}V
 =  
\int_{\partial V} \vec{r} (\vec{\nabla} \phi \cdot {\rm d}\vec{S})
- \int_V \vec{r} \Delta  \phi {\rm d}V
\label{green}
\end{equation}
With the help of $R_{ij} = |\vec{r}_j - \vec{r}_i|$ 
as the distance between points $i$ and $j$, and  $\vec{r}^\perp$ as
$\vec{r}$ projected onto the plane of the face $A_{ij}$ (for definition see
Fig.~\ref{FigVoronoiSketch} and Fig~\ref{Sketch}),
$\vec{r}$ becomes $\vec{r} =  \vec{r}^\perp + (\vec{r}_i - \vec{r}_j)/2  +
\vec{r}_i$,  and then 
the right hand side can be discretized for the Voronoi mesh as follows:
\begin{eqnarray}
\int_{\partial V} \vec{r} (\vec{\nabla} \phi \cdot {\rm d}\vec{S})
& = & \sum_{j \ne i}  \int_{A_{ij}} \vec{r} (\vec{\nabla} \phi \cdot {\rm d}\vec{S})  
\quad	\quad	- \int_{V_{i}} \vec{r} \Delta  \phi {\rm d}V\\
& = & \sum_{j \ne i}  \int_{A_{ij}} \left(\frac{1}{2} \vec{R}_{ij}+ \vec{r}^\perp\right)
(\vec{\nabla} \phi \cdot \vec{e}_{ij}) {\rm d}S 
\quad		+ \vec{r}_{i} \sum_{j \ne i}  \int_{A_{ij}}   (\vec{\nabla} \phi \cdot \vec{e}_{ij}) {\rm d}S
\quad	\quad	- \int_{V_{i}} \vec{r} \Delta  \phi {\rm d}V\\
& = & \sum_{j \ne i}  (\vec{\nabla} \phi \cdot \vec{R}_{ij}) A_{ij}\left( \frac{1}{2}
\vec{e}_{ij}+ \frac{1}{R_{ij} A_{ij}}\int_{A_{ij}}  \vec{r}^\perp {\rm d}S \right) 
\quad		+ \vec{r}_{i}   \int_{V_{i}}   \Delta \phi  {\rm d}V
\quad	\quad	- \int_{V_{i}} \vec{r} \Delta  \phi {\rm d}V\\
& = & \sum_{j \ne i}  (\vec{\nabla} \phi \cdot \vec{R}_{ij}) A_{ij}\left( \frac{1}{2} \vec{e}_{ij}+ \frac{\vec{c}_{ij}}{R_{ij}}  \right) 
	\quad \quad - \quad \quad \int_{V_{i}} \left(\vec{r}-\vec{r}_i\right) \Delta  \phi {\rm d}V\\
\end{eqnarray}
The second term vanishes for linear scalar fields. It is therefore
only a second order correction that becomes negligible for
sufficiently smooth fields if the
points lie near the centroids of their cells so that  $\int_V
\left(\vec{r}-\vec{r}_i\right) {\rm d}V$ .\\
Now we use $\vec{R}_{ij} \cdot \vec{\nabla} \phi=(\phi_j-\phi_i)$, so that we
obtain
for the gradient estimate $\left( \vec{\nabla} \phi \right)_i$ at point $i$
\begin{eqnarray}
\left( \vec{\nabla} \phi \right)_i & = &  \frac{1}{V_i} \sum_{j \ne i}  (\phi_j-\phi_i)  A_{ij}\left( \frac{1}{2} \vec{e}_{ij}+ \frac{\vec{c}_{ij}}{R_{ij}}  \right) \\
& = &  \frac{1}{V_i} \sum_{j \ne i}  A_{ij} \left[ (\phi_j-\phi_i)  \frac{\vec{c}_{ij}}{R_{ij}} + (\phi_j + \phi_i) \frac{\vec{e}_{ij}}{2}  \right] .
\end{eqnarray}
We note that application of this gradient estimate
to the Euler momentum equation in the form
\begin{equation}
m_i \ddot \vec{r}_i =  - V_i \, \vec{\nabla} P_i
\end{equation}
yields
\begin{equation}
m_i \ddot \vec{r}_i	 = - \sum_{j \ne i}    A_{ij}\left[(P_j - P_i) \frac{\vec{c}_{ij}}{R_{ij}} \: + \: (P_j + P_i)\: \frac{\vec{e}_{ij}}{2}\right]
\end{equation}
which is consistent with the expression derived directly from the
Lagrangian.

\subsection{Divergence and curl}
To estimate the divergence and curl of the velocity which we need for
the viscosity calculation of Section~\ref{Artificial_viscosity} we use
the same reasoning as in \ref{green}.
\begin{equation}
\int_{V_i} \left( \vec{r} \nabla \left(\nabla \vec{v} \right) \: + \:
\nabla \vec{v} \right) {\rm d}V   =   \int_{\partial V} \vec{r} (\nabla
\vec{v} \cdot {\rm d}\vec{S}) .
\end{equation}
Provided that $ \frac{1}{V_i} \int_{V_i} \nabla \times \left( \nabla
\times \vec{v}\right)$ vanishes for a linear field we define our 
estimator for the divergence operator as
\begin{equation}
(\nabla \cdot \vec{v} )_i = - \frac{1}{V_i} \sum_{j \ne i} A_{ij}
  \left[( \vec{v_j}- \vec{v_i}) \cdot \left( \frac{1}{2} \vec{e}_{ij}+
    \frac{\vec{c}_{ij}}{R_{ij}} \right) \right].
\end{equation}
Similarly, the curl estimator can be defined in the form
\begin{equation}
(\nabla \times \vec{B} )_i = - \frac{1}{V_i} \sum_{j \ne i} A_{ij}
  \left[( \vec{B_j}- \vec{B_i}) \times \left( \frac{1}{2} \vec{e}_{ij}+
    \frac{\vec{c}_{ij}}{R_{ij}} \right) \right],
\end{equation}
where $\vec{B}$ denotes some vector field.
Again applying Gauss' theorem to $\nabla \phi$ the Laplacian of a scalar function
$ \phi$ can be computed as
\begin{equation}
\int_V \Delta \phi \:{\rm d}V  =  
\sum_{j \ne i}  \int_{A_{ij}} (\vec{\nabla} \phi \cdot {\rm d} \vec{S})
  \simeq \sum_{j \ne i}  A_{ij} \frac{\phi_j - \phi_i}{R_{ij}}.
\end{equation}

\subsection{Accuracy of the gradient} \label{Test of accuracy}

To test the accuracy of the numerical gradient estimate, we assume a
quadratic model function $\phi(\vec{r})$ with constant gradient and
Hesse matrix, of the form
\begin{equation}
\phi(\vec{r})= \phi_0   +  \vec{A} \vec{r}   +    \frac{1}{2} \vec{r}^T \vec{B} \vec{r}.  
\end{equation}
For definiteness, we set $\vec{B} = b \vec{I}$, where $\vec{I}$ is the
identity matrix.
We then populate a box of unit length on a side with a set of points,
and evaluate the function $\phi(\vec{r})$ at the coordinates of each of
the points. After constructing the Voronoi tessellation for the point
set, we then estimate the local gradient for each cell based on
\begin{equation} 
(\nabla \phi)_i  =  \frac{1}{V_i} \sum_j  (\phi_j-\phi_i) \:
  A_{ij}\left( \frac{1}{2} \vec{e}_{ij}+ \frac{\vec{c}_{ij}}{R_{ij}}
  \right),
\end{equation}
and alternatively also based on a simpler version of this formula
where the terms proportional to $\vec{c}_{ij}$ are omitted, which
corresponds to the simplest version of a Green-Gauss gradient
estimate. We use three different point distributions with $4096$
points in a box of size unity.  First we use a (i) random Poisson
point distribution, a (ii) relaxed point distribution obtained from
the VPH scheme where each cell has the same volume (obtained from the
top of Fig.~\ref{FigSettling}), and (iii) a distribution relaxed with
PPO (obtained from the bottom of Fig.~\ref{FigSettling}) where in
addition very round cells were produced in which the majority of the
points lies close to the geometric centres of the cells. In all three
cases we compare the magnitude of the estimated gradient vector to the
magnitude of $\vec{A}$, and we plot the median of the relative error
as a function of $ b/|\vec{A}|$.  To exclude boundary effects, only
cells whose neighbours do not overlap with the box boundary are
considered in the measurement.

\begin{figure}
\begin{center}
  \includegraphics[width=0.42\textwidth]{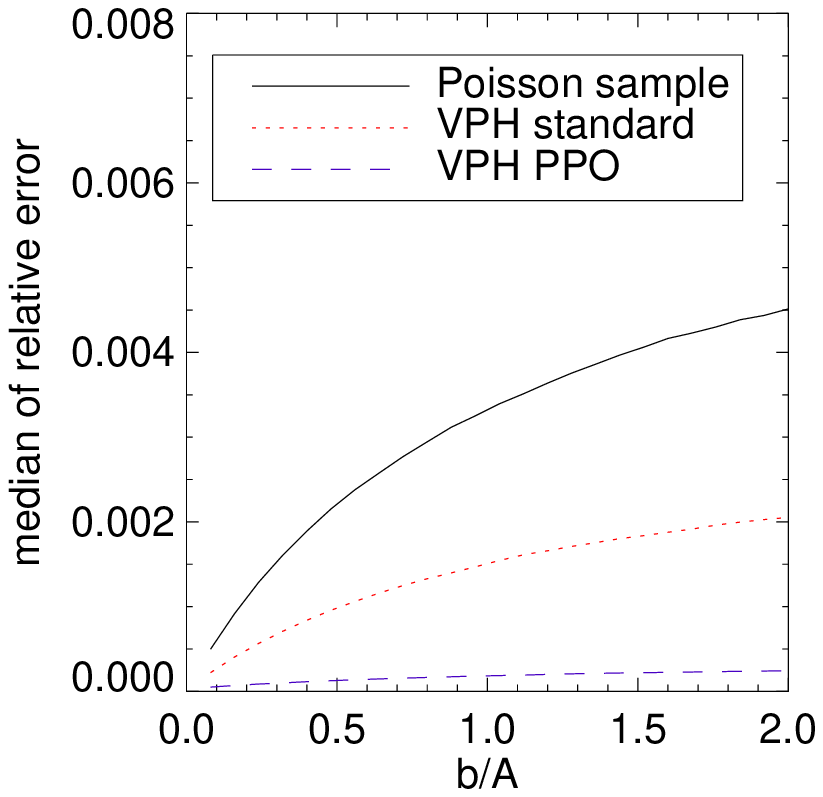}
  \includegraphics[width=0.42\textwidth]{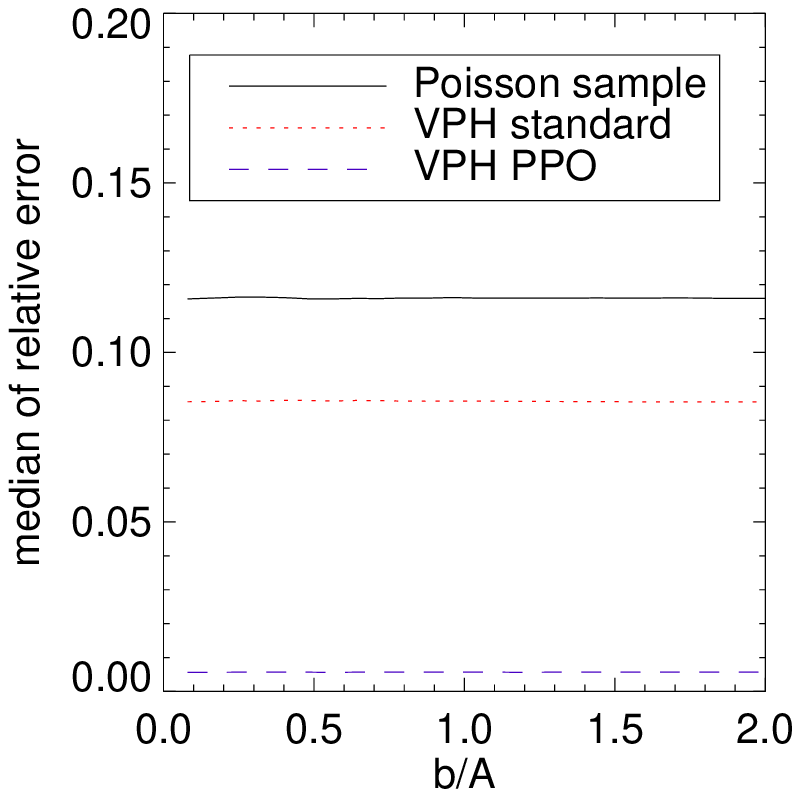}
\caption{Median relative error in the gradient estimate obtained
either with our default formula (left) or with the simpler Green-Gauss
estimate where the $\vec{c}_{ij}$ terms are omitted (right). We show
results for three different types of point sets, a Poisson sample
(black), a regularized distribution where each Voronoi cell has equal
volume (red), and a regularized distribution where in addition the
cells are quite `round' and regular (blue). The accuracy is measured as
function of the strength of a second order variation in the underlying field.
\label{FigGradAccuracy}
}
\end{center}
\end{figure}

In Figure~\ref{FigGradAccuracy}, we show the results. The panel on the
left gives our adopted gradient estimate, while the panel on the right
is for the simpler version of the Green-Gauss gradient
estimate. Interestingly, for $b=0$, the error vanishes exactly,
independent of the regularity of the Voronoi mesh. However, once the
second order term starts to influence the measurement, i.e.~for large
values if $ b/|\vec{A}|$, the more regular meshes clearly yield a lower
error, as expected. In all cases, the gradient estimate that includes
the $\vec{c}_{ij}$ term is superior to the simple Green-Gauss gradient
estimate. In particular, only when it is included, a vanishing error for
a linearly varying field is obtained.

\section{Controlling the shape of cells} \label{AppShape}

As outlined in Section~\ref{SecShapeCorrect}, we modify the fluid
Lagrangian slightly to include factors that penalize highly distorted
cell shapes. If such shapes occur, we want small adjustment forces to
appear that tend to make the mesh more regular again. These adjustment
forces need to preserve energy and momentum conservation of the scheme,
which will automatically be the case if they are derived from a suitably
defined Lagrangian or Hamiltonian. In this Appendix, we derive the 
equations of motion
for the Lagrangian
\begin{equation}
L =\sum_k \frac{1}{2} m_k\dot\vec{r}_k^2 -
\sum_k
\frac{P_kV_k}{\gamma-1}\left[1+\beta_0\frac{(\vec{r}_k-\vec{s}_k)^2}{V_k^{2/d}}\right]
\left\{ 
1+\beta_1 \left( \frac{\vec{w}_k^2}{V_k^{2/d}} -
\beta_2\right) 
\right\} ,
\end{equation}
where the factor in square brackets disfavours displacements of points
from the geometric centres of their cells, and the factor in curly
brackets disfavours cells with large aspect ratios.

We define the centroid of a cell as
\begin{equation}
\vec{s}_k \equiv \left<\vec{r}\right>_k = \frac{1}{V_k}\int 
\vec{r}\, \chi_k(\vec{r})\, {\rm d}\vec{r} ,
\end{equation}
where $\chi_k$ is the characteristic function of cell $k$. The shape
of a cell is measured via the second moment
\begin{equation}
\vec{w}_k^2 \equiv \left<(\vec{r}-\vec{s}_k)^2\right>_k = \frac{1}{V_k}\int
    (\vec{r}-\vec{s}_k)^2
\chi_k(\vec{r}) \, {\rm d}\vec{r}.
\end{equation}
Here $d$ counts the number of dimensions, i.e.~$d=2$ for 2D and $d=3$
for 3D. The factor $V_k^{2/d}$ is hence proportional to the `radius'
$R_k = V_k^{1/d}$ of a cell squared.  $\beta_0$ measures the strength
of the effect of displacements of points from the centroid of a cell,
while $\beta_1$ is the corresponding factor for the aspect-ratio
factor.  The constant $\beta_2$ is only introduced to prevent that
even round cells lead to a significant enhancement of the thermal
energy. For perfectly round cells, we expect in 2D roughly circles for which
$w_k^2 = V^{2/d} / (2\pi)$, hence we pick $\beta_2 = 1/(2\pi)$. In 3D,
we have approximately spheres instead and we pick $\beta_2 = 3/5
(3/4\pi)^{2/3}$.

Note that this Lagrangian is {\em only} a function of the point
coordinates for given entropies, so the equations of motion for
conservative dynamics are perfectly well defined, even though they lead
to more lengthy expressions than in the standard case. We first obtain
the following Lagrangian equation of motion:
\begin{equation}
m_i\ddot\vec{r}_i = \frac{\partial L}{\partial \vec{r}_i} = 
-\sum_k \frac{1}{\gamma-1} \left( \frac{\partial}{\partial \vec{r}_i}
P_k V_k\right)\Big[\, \Big] \Big\{\, \Big\}
\;-\; \sum_k \frac{P_k V_k}{\gamma-1} \Big\{\, \Big\}\beta_0
\frac{\partial}{\partial \vec{r}_i}
\frac{(\vec{r}_k-\vec{s}_k)^2}{V_k^{2/d}}
\;-\; \sum_k \frac{P_k V_k}{\gamma-1} \Big[\, \Big]\beta_1
\frac{\partial}{\partial \vec{r}_i}
\frac{w_k^2}{V_k^{2/d}} ,
\label{eqA1}
\end{equation}
where the empty square and curly brackets are notational short-cuts for
the corresponding terms in the original Lagrangian.  We note that we can use the identity
\begin{equation}
\frac{\partial}{\partial \vec{r}_i} P_k V_k = (1-\gamma) P_k
\frac{\partial V_k}{\partial \vec{r}_i}.
\end{equation}
We also note that in the second and third terms of equation (\ref{eqA1})
we encounter partial derivatives of $V_k$, which we can combine with the
first term into a more compact form. This allows us to write the
equation of motion in the form
\begin{equation}
m_i\ddot\vec{r}_i = \sum_k P_k^\star \frac{\partial V_k}{\partial  \vec{r}_i}
\;-\; \sum_k Q_k \frac{\partial}{\partial \vec{r}_i} (\vec{r}_k-\vec{s}_k)^2
\;-\; \sum_k L_k \frac{\partial}{\partial \vec{r}_i} w_k^2
\label{eqA2}
\end{equation}
where we have defined
\begin{equation}
P_k^\star \equiv P_k\left( \Big[\, \Big] \Big\{\, \Big\} 
+\frac{2}{d}\frac{\beta_0}{\gamma-1}
\frac{(\vec{r}_k-\vec{s}_k)^2}{V_k^{2/d}} \Big\{\, \Big\}
 +\frac{2}{d}\frac{\beta_1}{\gamma-1}
\frac{w_k^2}{V_k^{2/d}} \Big[\, \Big]
\right)
\end{equation}
and introduced the quantities
\begin{equation}
Q_k\equiv \frac{\beta_0}{\gamma-1} P_k V_k^{1-2/d}  \Big\{\, \Big\},
\end{equation}
\begin{equation}
L_k\equiv \frac{\beta_1}{\gamma-1} P_k V_k^{1-2/d}  \Big[\, \Big].
\end{equation}
We already know an explicit expression for ${\partial V_k}/{\partial
  \vec{r}_i}$, but we still need to derive such a thing for the derivatives
of $(\vec{r}_k-\vec{s}_k)^2$ and $w_k^2$. Let us first deal with the
term involving $Q_k$ in the equations of motion, i.e.
\begin{equation}
\left(m_i\ddot\vec{r}_i\right)_{Q} =- \sum_k Q_k
\frac{\partial}{\partial \vec{r}_i} (\vec{r}_k-\vec{s}_k)^2
= -2 \sum_k Q_k \left(\frac{\partial(\vec{r}_k-\vec{s}_k)}{\partial
  \vec{r}_i}\right)^T (\vec{r}_k-\vec{s}_k).
\label{eqA3}
\end{equation}
Here the exponent $T$ stands for the transpose, and the notation
$\frac{\partial\vec{a}}{\partial \vec{b}}$ is the Jacobian matrix with
elements $\left(\frac{\partial\vec{a}}{\partial \vec{b}}\right)_{lm}
= \frac{\partial a_l}{\partial b_m}$. Based on the definition of
$\vec{s}_k$ in terms of the characteristic function we find
\begin{equation}
\frac{\partial}{\partial \vec{r}_i} (\vec{r}_k-\vec{s}_k)
= \delta_{ki} \mathbf{1} -\frac{1}{V_k}(\vec{r}_k-\vec{s}_k)
\left(\frac{\partial V_k}{\partial \vec{r}_i}\right)^T
+\frac{1}{V_k} \int {\rm d} \vec{r} \, ( \vec{r}_k-\vec{r})
  \left(\frac{\partial \chi_k}{\partial \vec{r}_i}\right)^T .
\end{equation}
For the derivative of the characteristic function we can use a result
from
\citet{Serrano} and
write
\begin{equation}
\frac{\partial \chi_k}{\partial \vec{r}_i} = 
\sum_j \delta_{ki} \frac{ \chi_k\chi_j}{\sigma^2} \,  (\vec{r}-\vec{r}_k)
- \frac{\chi_k\chi_i}{\sigma^2} \,(\vec{r}-\vec{r}_i),
\end{equation}
which is based on approximating the characteristic function with
\begin{equation}
\chi_k(\vec{r}) = 
\frac{\exp\left[-\frac{(\vec{r}-\vec{r}_k)^2}{2\sigma^2}\right]}
{\sum_j\exp\left[-\frac{(\vec{r}-\vec{r}_j)^2}{2\sigma^2}\right]} ,
\end{equation}
which becomes exact in the limit $\sigma \to 0$.
Putting these results into equation~(\ref{eqA3}) one gets
\begin{eqnarray}
\left(m_i\ddot\vec{r}_i\right)_{Q}
& = &  - 2 Q_i (\vec{r}_i - \vec{s}_i) + 
2 \sum_k \frac{Q_k}{V_k} (\vec{r}_k-\vec{s}_k)^2 \frac{\partial
  V_k}{\partial \vec{r}_i}
-2\sum_j \frac{Q_i}{V_i} \int {\rm d}\vec{r}
\frac{\chi_i \chi_j}{\sigma^2} (\vec{r}-\vec{r}_i)
(\vec{r}_i-\vec{r})^T(\vec{r}_i-\vec{s}_i) \nonumber \\
& & +2\sum_k \frac{Q_k}{V_k} \int {\rm d}\vec{r}
\frac{\chi_i \chi_k}{\sigma^2} (\vec{r}-\vec{r}_i) (\vec{r}_k-\vec{r})^T(\vec{r}_k-\vec{s}_k).
\label{eqA5}
\end{eqnarray}
We can now identify the area of a face between two cells as 
\begin{equation}
A_{ij} = R_{ij} \int   {\rm d}\vec{r}
\frac{\chi_i \chi_j}{\sigma^2},
\end{equation}
and the centroid of the face as 
\begin{equation}
\vec{s}_{ij} = \frac{R_{ij}}{A_{ij}}  \int   {\rm d}\vec{r}
\frac{\chi_i \chi_j}{\sigma^2} \vec{r}.
\end{equation}
Furthermore, we define a second-order tensor of the face relative its
centroid as 
\begin{equation}
\vec{T}_{ij} = \frac{R_{ij}}{A_{ij}}  \int   {\rm d}\vec{r}
\frac{\chi_i \chi_j}{\sigma^2} (\vec{r}-\vec{s}_{ij})(\vec{r}-\vec{s}_{ij})^T.
\end{equation}
With these definitions, we can rewrite equation (\ref{eqA5}) as 
\begin{equation}
\left(m_i\ddot\vec{r}_i\right)_{Q} =  - 2 Q_i (\vec{r}_i - \vec{s}_i) + 
2 \sum_k \frac{Q_k}{V_k} (\vec{r}_k-\vec{s}_k)^2 \frac{\partial
  V_k}{\partial \vec{r}_i}
+2\sum_{j\ne i} \frac{A_{ii}}{R_{ij}}
\left\{  \vec{T}_{ij}(\vec{e}_i-\vec{e}_j) +
\left[(\vec{s}_{ij}-\vec{r}_i)\vec{e}_i  -
  (\vec{s}_{ij}-\vec{r}_j)\vec{e}_j \right] (\vec{s}_{ij}-\vec{r}_i) \right\},
\end{equation}
where we introduced the further short-cut
\begin{equation}
\vec{e}_i \equiv \frac{Q_i}{V_i} (\vec{r}_i-\vec{s}_i) .
\end{equation}
We note that the second term in this equation can be absorbed in yet a
further redefinition of $P_k^\star$, which we will exploit later on.
We next consider the term in the
full equation of motion that involves the $L_k$ factor. This is given by
\begin{eqnarray}
\left(m_i\ddot\vec{r}_i\right)_{L} & = & - \sum_k L_k
\frac{\partial w_k^2}{\partial \vec{r}_i}
 =- \sum_k L_k \left[ -\frac{w_k^2}{V_k}\frac{\partial V_k}{\partial
     \vec{r}_i} + \frac{1}{V_k} 
\int {\rm d}\vec{r} \frac{\partial \chi_k}{\partial \vec{r}_i}
(\vec{r}-\vec{s}_k)^T (\vec{r}-\vec{s}_k) \right]  \label{eqA9}
 \\
& =  & 
\sum_k \frac{L_k w_k^2}{V_k}\frac{\partial V_k}{\partial
     \vec{r}_i} 
-
\sum_j \frac{L_i}{V_i}\int {\rm d}\vec{r} \frac{\chi_i
  \chi_j}{\sigma^2}
(\vec{r}-\vec{r}_i) (\vec{r}-\vec{s}_i)^T (\vec{r}-\vec{s}_i)
+
\sum_j \frac{L_j}{V_j}\int {\rm d}\vec{r} \frac{\chi_i
  \chi_j}{\sigma^2}
(\vec{r}-\vec{r}_i) (\vec{r}-\vec{s}_j)^T (\vec{r}-\vec{s}_j).
\nonumber
\end{eqnarray}
We now define a further moment for each cell face, namely the
vector-valued quantity
\begin{equation}
\vec{g}_{ij} \equiv  \frac{R_{ij}}{A_{ij}} \int {\rm d}\vec{r} \frac{\chi_i
  \chi_j}{\sigma^2} (\vec{r}-\vec{s}_{ij})^2\,(\vec{r}-\vec{s}_{ij}).
\end{equation}
Note that $\vec{g}_{ij}$ always vanishes in 2D but can be non-zero in
3D. With this definition, we can rewrite equation (\ref{eqA9}) as
\begin{eqnarray}
\left(m_i\ddot\vec{r}_i\right)_{L} & = &
\sum_k \frac{L_k w_k^2}{V_k}\frac{\partial V_k}{\partial
     \vec{r}_i} \\
& + &
\sum_{j \ne i} \frac{A_{ij}}{R_{ij}} 
\left\{
\left( \frac{L_j}{V_j}-\frac{L_i}{V_i}\right)
\vec{g}_{ij}
+2 \vec{T}_{ij} (\vec{f}_j -\vec{f}_i) +
\left[
(\vec{f}_j (\vec{s}_{ij}-\vec{s}_j) -
\vec{f}_i(\vec{s}_{ij}-\vec{s}_i))
+ Tr(\vec{T}_{ij}) \left(\frac{L_j}{V_j}-\frac{L_i}{V_i}\right)
\right] (\vec{s}_{ij} - \vec{r}_i) ,
\right\}\nonumber
\end{eqnarray}
where we have defined the short-cut
\begin{equation}
\vec{f}_i\equiv \frac{L_i}{V_i} (\vec{s}_{ij} - \vec{s}_i) .
\end{equation}
Again, the first term involving $\partial V_k / \partial \vec{r}_i$
can be absorbed into a redefinition of $P_k^\star$. Putting everything
together, the complete equation of motion can then be written as
\begin{eqnarray}
m_i\ddot\vec{r}_i  & = &
\sum_k P_k^{\star \star} \frac{\partial V_k}{\partial \vec{r}_i}
\; - 2 Q_i (\vec{r}_i - \vec{s}_i)\nonumber \\
& + &
\sum_{j \ne i} \frac{A_{ij}}{R_{ij}} 
\bigg\{
2\vec{T}_{ij} (\vec{e}_i -\vec{e}_j +  \vec{f}_j -\vec{f}_i)\nonumber
\\
& & \;\;\;\;\;\;\;\;\;\;\;\;\;\; + \left[
2(\vec{s}_{ij}-\vec{r}_i)\vec{e}_i -
2(\vec{s}_{ij}-\vec{r}_j)\vec{e}_j
+ (\vec{s}_{ij}-\vec{s}_j)\vec{f}_j -
(\vec{s}_{ij}-\vec{s}_i)\vec{f}_i    + Tr(\vec{T}_{ij})
\left(\frac{L_j}{V_j}-\frac{L_i}{V_i}\right)
\right] (\vec{s}_{ij}-\vec{r}_i) \nonumber \\
& &\;\;\;\;\;\;\;\;\;\;\;\;\;\; + \left(\frac{L_j}{V_j}-\frac{L_i}{V_i}\right)
\vec{g}_{ij} \bigg\} ,
\end{eqnarray}
where we have defined
\begin{equation}
P_k^{\star\star} = P_k^{\star} + \frac{L_k w_k^2}{V_k} + \frac{2Q_k
  (\vec{r}_k-\vec{s}_k)^2}{V_k} .
\end{equation}
While a bit lengthy, this can be straightforwardly calculated for the
VPH scheme. Nevertheless, we want to add a brief note on how to compute
the Tensors $T_{ij}$, which is done as part of the mesh construction. We
have
\begin{equation}
\left<(\vec{r}-\vec{s})^2\right>_k =
\left<(\vec{r}-\vec{r}_0)^2\right>_k - (\vec{r}_0 - \vec{s})^2
\end{equation}
for any reference point $\vec{r}_0$. Suppose we have a triangle in 2D given by
$(\vec{r}_0, \vec{r}_1, \vec{r}_2$), then the moment can be obtained as
\begin{equation}
\left<(\vec{r}-\vec{r}_0)^2\right>_k = \frac{1}{6}
\left[ (\vec{r}_1 - \vec{r}_0)  (\vec{r}_1 - \vec{r}_0)^T
+ (\vec{r}_1 - \vec{r}_0)  (\vec{r}_2 - \vec{r}_0)^T
+ (\vec{r}_2 - \vec{r}_0)  (\vec{r}_2 - \vec{r}_0)^T \right].
\end{equation}
Similar relations hold for 3D and can be exploited for an efficient
calculation of the tensors $\vec{T}_{ij}$ and the vectors
$\vec{g}_{ij}$.

\label{lastpage}

\end{document}